\documentclass[12pt]{article}
\usepackage{latexsym}
\usepackage{amsmath}
\usepackage{amssymb}
\catcode`\@=11
\newcount\hour
\newcount\minute
\newtoks\amorpm \hour=\time\divide\hour by 60\minute
=\time{\multiply\hour by 60 \global\advance\minute by-\hour}
\edef\standardtime{{\ifnum\hour<12 \global\amorpm={am}%
        \else\global\amorpm={pm}\advance\hour by-12 \fi
        \ifnum\hour=0 \hour=12 \fi
        \number\hour:\ifnum\minute<10
        0\fi\number\minute\the\amorpm}}
\edef\militarytime{\number\hour:\ifnum\minute<10
0\fi\number\minute}
\def\draftlabel#1{{\@bsphack\if@filesw {\let\thepage\relax
   \xdef\@gtempa{\write\@auxout{\string
      \newlabel{#1}{{\@currentlabel}{\thepage}}}}}\@gtempa
   \if@nobreak \ifvmode\nobreak\fi\fi\fi\@esphack}
        \gdef\@eqnlabel{#1}}
\def\@eqnlabel{}
\def\@vacuum{}
\def\marginnote#1{}
\def\draftmarginnote#1{\marginpar{\raggedright\scriptsize\tt#1}}
\def\draft{
        \pagestyle{plain}
        \overfullrule=2pt
        \oddsidemargin -.1truein
        \def\@oddhead{\sl \phantom{\today\quad\militarytime} \hfil
        \smash{\Large\sl DRAFT} \hfil \today\quad\militarytime}
        \let\@evenhead\@oddhead
        \let\label=\draftlabel
        \let\marginnote=\draftmarginnote
        \def\ps@empty{\let\@mkboth\@gobbletwo
        \def\@oddfoot{\hfil \smash{\Large\sl DRAFT} \hfil}
        \let\@evenfoot\@oddhead}
        \def\@eqnnum{(\theequation)\rlap{\kern\marginparsep\tt\@eqnlabel}%
        \global\let\@eqnlabel\@vacuum}  }

\renewcommand{\theequation}{\thesection.\arabic{equation}}
\renewcommand{\thefootnote}{\fnsymbol{footnote}}
\newcommand{\newsection}{    
\setcounter{equation}{0}\section}
\def\appendix#1{\addtocounter{section}{1}\setcounter{equation}{0}
\renewcommand{\thesection}{\Alph{section}}
\section*{Appendix \thesection\protect\indent \parbox[t]{11.15cm}{#1}}
\addcontentsline{toc}{section}{Appendix \thesection\ \ \ #1}}

\def \la {\label}

\def \b {\beta}

\def\be{\begin{equation}}
\def\ee{\end{equation}}

\catcode`\@=11

\def\bea{\begin{eqnarray}}
\def\eea{\end{eqnarray}}
\def\beann{\begin{eqnarray*}}
\def\eeann{\end{eqnarray*}}
\def\beq{\begin{equation}}
\def\eeq{\end{equation}}
\def\ba{\begin{array}}
\def\ea{\end{array}}
\def\ben{\begin{enumerate}}
\def\een{\end{enumerate}}

 \def \la {\label}
 \def\be{\begin{equation}}
\def\ee{\end{equation}}

\def \la {\label}

\font\mybb=msbm10 at 11pt

\def\bb#1{\hbox{\mybb#1}}

\def\bZ {\bb{Z}}
\def\bR {\bb{R}}

\def\bC {\bb{C}}

\def\e  {\epsilon}

\def \ee {\epsilon}

\def\a{\alpha }

\def \b {\beta}

\def\be{\begin{equation}}
\def\ee{\end{equation}}

\def \la{\label}

\begin{document}
\date{November 2002}
\begin{titlepage}
\begin{center}
\vspace*{-1.0cm}
\hfill UB-ECM-PF-07-28 \\

\vspace{2.0cm} {\Large \bf  IIB solutions with N$>$28 Killing spinors are maximally supersymmetric} \\[.2cm]

\vspace{1.5cm}
 {\large  U. Gran$^1$, J. Gutowski$^2$,  G. Papadopoulos$^3$ and D. Roest$^4$}

\vspace{0.5cm}

${}^1$ Fundamental Physics\\
Chalmers University of Technology\\
SE-412 96 G\"oteborg, Sweden\\

\vspace{0.5cm}
${}^2$ DAMTP, Centre for Mathematical Sciences\\
University of Cambridge\\
Wilberforce Road, Cambridge, CB3 0WA, UK

\vspace{0.5cm}
${}^3$ Department of Mathematics\\
King's College London\\
Strand\\
London WC2R 2LS, UK\\

\vspace{0.5cm}
${}^4$ Departament Estructura i Constituents de la Materia \\
    Facultat de F\'{i}sica, Universitat de Barcelona \\
    Diagonal, 647, 08028 Barcelona, Spain \\

\end{center}

\vskip 1.5 cm
\begin{abstract}
\noindent  We show that all IIB supergravity backgrounds which admit more than 28 Killing spinors
are maximally supersymmetric. In particular, we find that for all $N>28$ backgrounds the supercovariant curvature
vanishes, and that the
 quotients of maximally supersymmetric backgrounds either preserve all 32  or  $N<29$ supersymmetries.

\end{abstract}

\end{titlepage}
\newpage
\setcounter{page}{1}
\renewcommand{\thefootnote}{\arabic{footnote}}
\setcounter{footnote}{0}

\setcounter{section}{0}
\setcounter{subsection}{0}
\newsection{Introduction}

Recently, it has been realized that there are restrictions on the existence of type II and eleven-dimensional
supergravity backgrounds with near maximal number of supersymmetries. This was initiated in \cite{N=31-IIB} where it was shown that IIB
backgrounds with $N=31$ supersymmetries are maximally supersymmetric. Later this was extended to
IIA backgrounds in \cite{N=31-IIA}. These results mostly follow from an analysis of the algebraic Killing spinor equations.

Eleven-dimensional supergravity backgrounds with $31$ supersymmetries also admit an additional
Killing spinor and so are maximally supersymmetric. To show this, one first proves  that
the supercovariant curvature of $N=31$ backgrounds vanishes subject to the field equations and Bianchi identities
of eleven-dimensional supergravity \cite{N=31-M}. This demonstrates that the $N=31$ backgrounds are locally maximally supersymmetric.
Then one shows that there are  no discrete quotients of maximally supersymmetric backgrounds which preserve
 $31$ supersymmetries \cite{N=31-M2}. These results exclude the existence of preonic  backgrounds  \cite{bandos} in type II and eleven-dimensional
supergravities.

Most of the above results have been obtained  by adapting the spinorial geometry technique for solving
Killing spinor equations  \cite{uggp}
to backgrounds with near maximal number of
supersymmetries. The investigation of discrete quotients of maximally supersymmetric
backgrounds relies on techniques developed in \cite{simona, simonb}.
Similar results hold for some supergravities in lower dimensions \cite{gutsabra}. However in non-maximal
supergravities in four and five dimensions, it is possible to construct preonic backgrounds as discrete quotients
of maximally supersymmetric ones \cite{janjose}.

In this paper, we  show  that IIB backgrounds with $N>28$ supersymmetries are maximally supersymmetric.
For this, we first use the property that $N>24$ supersymmetric IIB backgrounds
are homogeneous spaces  \cite{N>24}. This
in particular implies that the one-form field strength $P$ vanishes, $P=0$. As a result
 the algebraic Killing spinor equation of IIB supergravity is linear over the complex numbers and so
it always has an even number of solutions. In addition, an application of the spinorial geometry technique reveals
that if $N=30$, then the three-form field strength vanishes as well, $G=0$. Therefore  one  concludes
that for all $N>28$ IIB backgrounds, the algebraic Killing spinor equation implies $P=G=0$.

This in turn implies
that  the gravitino Killing spinor equation also has even number of solutions \cite{N=31-IIB}.
Therefore
to prove our result, we should exclude the existence of IIB backgrounds with 30 supersymmetries.
For this we explore the integrability conditions of the gravitino Killing spinor equation. The analysis is
similar in spirit as that for the $N=31$ backgrounds of eleven-dimensional supergravity \cite{N=31-M}. In particular,
we show that the curvature ${\cal R}$ of the supercovariant connection vanishes, ${\cal R}=0$, subject to
the Bianchi identities and field equations of IIB supergravity. This demonstrates that
$N>28$ IIB backgrounds are locally  maximally supersymmetric.  Using the classification
of maximally supersymmetric IIB
backgrounds  \cite{gpjose}, one concludes that the $N>28$ backgrounds must be locally isometric to one of the following solutions:  Minkowski
space $\bR^{9,1}$,  the Freund-Rubin space $AdS_5\times S^5$ \cite{schwarz} and the maximally supersymmetric
plane wave \cite{mspw}.

Finally, we show that one cannot construct $28<N<32$
IIB backgrounds as discrete quotients of the  maximally supersymmetric ones. To establish our result, we lift the
generators of the discrete symmetry group to $Spin_c(9,1)=Spin(9,1)\times_{\bZ_2} U(1)$ and prove that there are no invariant spinors
that span a 30-dimensional subspace. This computation relies
 on the lift of the generators of the discrete group to the $Spin(9,1)$ group investigated in \cite{simona, simonb}. Our
lift has an additional phase along the $U(1)$ direction of $Spin_c(9,1)$.

This paper is organized as follows. In section two, we show using the algebraic Killing spinor equation
that for $N>28$ supersymmetric IIB
backgrounds the three-form field strength vanishes, $G=0$. In section three, we describe the
conditions that the field equations and the Bianchi identities impose on the holonomy
of the supercovariant IIB connection. In sections four, five and six, we demonstrate that the supercovariant
curvature of all $N>28$ IIB backgrounds vanishes. In section seven, we exclude the possibility
of constructing $28<N<32$ backgrounds as discrete quotients of Minkowski space $\bR^{9,1}$, $AdS_5\times S^5$ and the
maximally supersymmetric plane wave, and in section eight we give our conclusions.

\newsection{Algebraic Killing spinor equation}

The algebraic Killing spinor equation (KSE) of IIB supergravity \cite{ws, schwarz, howe} is
 \bea
  P_A \Gamma^A C \epsilon^* + {1\over24} G_{ABC}\Gamma^{ABC} \epsilon = 0 ~,
 \eea
 where $P$ and $G$ are the (complex) one- and three-form field strengths, respectively, $C$ is the charge
 conjugation matrix, and $\e$ is a complex Weyl $Spin_c(9,1)$ spinor. For our spinor conventions, see e.g.~\cite{ugjggp}.
It is known that IIB backgrounds with more than 24 supersymmetries
are locally homogeneous \cite{N>24}. In particular, this implies that the
scalars are constant and hence that their field strength vanishes, $P=0$. The
vanishing of $P$ has the important implication that the dilatino KSE
becomes linear over the complex numbers. In other words, it has an
even number of solutions which can be expressed as  $(\epsilon^r, i
\epsilon^r)$ pairs.

The aim is to show that the algebraic Killing spinor equation for $N>28$ backgrounds implies
$G=0$. It is known that if $N=32$, the algebraic Killing spinor equation implies that $P=G=0$
\cite{gpjose}. So it remains to prove the statement for $N=30$. Since the algebraic Killing spinor
equation for $P=0$ is linear over the complex numbers, the solution spans a complex hyperplane
 in the space of spinors at every spacetime point.
So it  has a  normal $\nu$ with respect to the standard Majorana inner product.
Using spinorial geometry and in particular the gauge symmetry of the
Killing spinor equations, the normal direction $\nu$ can be chosen of the form \cite{N=31-IIB}
 \bea
 Spin(7) \ltimes \bR^8: && \nu =(n+im) (e_5+e_{12345}) \,, \cr
 SU(4) \ltimes \bR^8: && \nu =(n-\ell+im) e_5+ (n+\ell+im) e_{12345} \,, \cr
 G_2: && \nu = n(e_5+e_{12345})+i m (e_1+e_{234}) \,, \label{normals}
 \eea
corresponding to the three different orbits of $Spin(9,1)$ in the space of negative chirality Weyl spinors \cite{ugjggp}, where
$n,m$ and $\ell$ are real spacetime functions. Choosing the solutions orthogonal to the above normals, they can be expressed
as
\bea
\e^r=\sum_{s=1}^{15} z^r{}_s\eta^s~,
\eea
where $\eta_i$ is a basis normal to $\nu$ and $z$ is an invertible $15\times 15$ matrix of spacetime dependent complex functions,
see \cite{iibsyst} for more details.
Consequently, the Killing spinor equation becomes
\bea
G_{ABC}\Gamma^{ABC}\eta^r=0~.
\la{akse}
\eea
Since in all three cases the normal  $\nu$ can be arranged to  point only in at most three different directions
$e_5 + e_{12345}, i (e_5-e_{12345})$ and $(e_1+e_{234})$, the bases $(\eta^s)$ can be chosen such that they contain  13
common elements. The other two elements depend on the choice of orbit  and
have to be considered case by case.
 We will first analyze the constraints obtained from ({\ref{akse}})
 acting on the 13 common elements, and afterwards
specialize to the three different cases.

The 13 common basis elements $\eta^r$, $r=1,\ldots,13$, are given by those of the 16 basis elements of the Majorana-Weyl representation
of $Spin(9,1)$ which are linearly independent from  $1+e_{1234}, i(1-e_{1234})$
and $(e_{15} + e_{2345})$. Substituting this into the algebraic Killing spinor equation (\ref{akse}), we find that
the non-vanishing components of $G$ satisfy
 \begin{align}
  & G_{m \bar 1 \bar m} = - \tfrac12 G_{\bar 2 \bar 3 \bar 4} \,, \quad
  G_{- + \bar 1} = \tfrac12 G_{\bar 2 \bar 3 \bar 4} \,, \quad
  G_{+ 1 \bar 1} = G_{+ m \bar m} \,, \notag \\
  & G_{1 m \bar m} = - \tfrac12 G_{234} \,, \quad
  G_{- + 1} = \tfrac12 G_{234} \,,
 \end{align}
where $m = 2,3,4$, and there is no summation in the repeated $m$ indices. Hence there are only three independent non-vanishing components left of the original 120.

Now the analysis splits up for the three different orbits, since the two
additional basis elements  $\eta^r$, $r=14,15$, differ:
 \begin{itemize}
  \item
  The simplest orbit is $Spin(7) \ltimes \bR^8$, in which case the two
additional  basis elements are $\eta^{14} = 1 - e_{1234}$ and $\eta^{15} = e_{15}
+ e_{2345}$. When inserted into the dilatino variation, the former implies $G_{+ 1 \bar 1} = 0$ and the latter
implies $G_{234} = G_{\bar 2 \bar 3 \bar 4} = 0$. Hence $G=0$ in this case.
 \item
In the $SU(4) \ltimes \bR^8$ case, one  has $\eta^{14} = e_{15} + e_{2345}$. This leads to  $G_{234} = G_{\bar 2 \bar 3 \bar 4} =
0$. The remaining basis element is given by $\eta^{15} = (n - \ell + im) 1 - (n + \ell + im) e_{1234}$
and implies $G_{+ 1 \bar 1} = 0$. Hence $G$ also vanishes for the $SU(4) \ltimes \bR^8$ orbit.
 \item
The remaining case is the $G_2$ orbit. For this,  $\eta^{14} = 1 -
e_{1234}$, which leads to the vanishing of $G_{+ 1 \bar 1}$. The other two components of $G$ are set
to zero by  $\eta^{15} = m (1 + e_{1234}) + i n (e_{15} + e_{2345})$.
Hence $G=0$ for this orbit as well.
 \end{itemize}

Therefore we conclude that for $N>28$ IIB backgrounds, $P=G=0$ as a consequence of  the homogeneity
and the algebraic Killing spinor equation. As we have mentioned, if $G=0$, the gravitino Killing spinor equation
has an even number of solutions. Thus $N>28$ IIB backgrounds can have either 30 or 32 supersymmetries.
We shall exclude the existence of $N=30$ backgrounds by investigating the gravitino Killing spinor equation.

\newsection{ Supercovariant curvature and holonomy}

\subsection{Supercurvature}

Assuming $G=0$, the curvature ${\cal R}=[{\cal D}, {\cal D}]$ of the covariant connection ${\cal D}$
 of IIB supergravity can be expanded \cite{gpjose}  as
\bea
{\cal R}_{MN}={\rm Re}\,{\cal R}_{MN}+i {\rm Im}\, {\cal R}_{MN}= {1\over 2} (T_{MN}^2)_{PQ} \Gamma^{PQ}
+{1\over 4!} (\hat T^4_{MN}+ i \tilde T^4_{MN})_{Q_1\dots Q_4} \Gamma^{Q_1\dots Q_4}~,
\label{Rexpansion}
\eea
where
\bea
(T^2_{MN})_{P_1P_2} &=& \tfrac{1}{4} R_{M N, P_1P_2}-\tfrac{1}{12}F_{M[P_1}{}^{Q_1Q_2Q_3}F_{|N|P_2]Q_1Q_2Q_3}~,\notag\\
(T^4_{MN})_{P_1 \ldots P_4} &=& \tfrac{i}{2}D_{[M}F_{N]P_1\ldots P_4}+\tfrac{1}{2}F_{M N Q_1Q_2 [P_1}F_{P_2 P_3 P_4]}{}^{Q_1 Q_2}~,
\label{Tphys}
\eea
and $R$ is the Riemann curvature, $F$ is the self-dual five-form field strength and $T^4 = \hat T^4 + i \tilde T^4$.
Observe that $\tilde T^4$  contains only the covariant derivative
of $F$.
We have made use of the self-duality of $F$ to simplify these
expressions. The components of $T^2$ and $T^4$ are not all independent but are restricted by the Bianchi identities
of $R$ and $F$, ($dF=0$), and the field equations of IIB supergravity. In particular, using the expressions of $T^2$ and $T^4$
in terms of the physical fields (\ref{Tphys}) and the Bianchi identities, one finds that
\bea
(T^2_{MN})_{P_1P_2} &=& (T^2_{P_1P_2})_{MN}~,\notag\\
(T^2_{M[P_1})_{P_2P_3]} &=& 0~,\notag\\
(T^4_{[P_1P_2})_{P_3 P_4 P_5 P_6]} &=& 0~.
\label{physcond}
\eea
Next observe that $\Gamma^N {\cal R}_{MN}$ is a linear combination of the field equations \cite{iibsyst}. Making use of this and
of (\ref{physcond}), we find
\bea
(T^2_{MN})_{P}{}^N &=& 0 ~,\notag\\
(T^4_{MN})_{P_1P_2P_3}{}^N &=& 0 ~,\notag\\
(T^4_{M[P_1})_{P_2 P_3 P_4 P_5]} &=& -{1 \over 5!}
\epsilon_{P_1 P_2 P_3 P_4 P_5}{}^{Q_1 Q_2 Q_3 Q_4 Q_5}
(T^4_{M[Q_1})_{Q_2 Q_3 Q_4 Q_5]}~.
\label{feqcond}
\eea
Also note that $(T^4{}_{P_1 (M})_{N) P_2 P_3 P_4}$ is totally antisymmetric in $P_1$, $P_2$, $P_3$, $P_4$.

One of the consequences of the first condition in (\ref{feqcond}), or equivalently from the
Einstein field equation and $P=G=0$, is that the scalar curvature of the spacetime vanishes, i.e.~$R=0$.
Furthermore, on imposing the Einstein equations, and using the
self-duality of $F$, it is straightforward to show that
$(T^2_{MN})_{PQ} = {1 \over 4} W_{MNPQ}$,
where $W$ is the spacetime Weyl tensor.
The expressions in this subsection do not rely on the existence of Killing spinors and are therefore valid for all backgrounds.

\subsection{Holonomy}

It is clear from the expression for ${\cal R}$ in the previous section that the (reduced) holonomy of the supercovariant connection of
 IIB backgrounds with
$P=G=0$ is contained in $SL(16, \bC)$. This is a subgroup of $SL(32,\bR)$ which is the holonomy
of the supercovariant connection for generic IIB backgrounds \cite{gptsimpis}. It immediately follows from the integrability conditions
of the gravitino Killing spinor equation and in particular of
\bea
{\cal R}\epsilon^r=0
\la{intcon}
\eea
that the holonomy of a spacetime
with $N=2n$ supersymmetries reduces to a subgroup of $SL(16-n, \bC)\ltimes \oplus_n \bC^{16-n}$. Therefore on the
grounds of holonomy, one expects that there are supersymmetric $P=G=0$ backgrounds with any even number $N\leq 32$ of
supersymmetries. However as we shall show the $N=30$ case will be excluded.

Let $(\e^r, \tilde \e^p)$ be a complex (local) basis in the space of  spinors where $\e^r$, $r=1,\dots,n$ is a basis
in the space of Killing spinors, $N=2n$, and $p=n+1,\dots, 16$. Moreover, let $\nu^q$, $q=1, \dots 16-n$, be a basis
in the space normal to the Killing spinors with respect to the Majorana inner product $B$.
Using a similar argument to the one we have employed for M-theory \cite{N=31-M}, the supercurvature
of a spacetime with $N=2n$ Killing spinors can be locally written as
\bea
{\cal R}_{MN, ab'}=U_{MN,rq} \epsilon_a^r \nu^q_{b'}+U_{MN,pq}\tilde\epsilon_a^p\nu^q_{b'}~,
\eea
where $a, b'$ are chiral and anti-chiral spinorial indices, respectively, and $U_{MN,rq}$ and $U_{MN,pq}$ are  complex spacetime
 two-forms. Clearly, in writing the supercovariant curvature in this way it automatically satisfies the
 integrability condition (\ref{intcon}).
Moreover, the above condition  can be written in any other basis in the space of spinors. In particular, we may choose
say a Majorana or another suitable basis $\eta^r$ and write
\bea
{\cal R}_{MN, ab'}=u_{MN,rq} \eta^r_{a} \nu^q_{b'}~,
\eea
where again $u$ are complex two-forms on the spacetime.
On the other hand we know that
\bea
\eta_a \theta_{b'}=-{1\over16} \sum_{k=0}^2 {1\over (2k)!} B(\eta, \Gamma_{A_1A_2\dots A_{2k}}\theta)
(\Gamma^{A_1A_2\dots A_{2k}})_{ab'}~,
\eea
This in turn gives
\bea
{\cal R}_{MN,A_1\dots A_{2k}}=-{1\over16} u_{MN,rq} B(\eta^r, \Gamma_{A_1A_2\dots A_{2k}}\nu^q)~.
\eea

The complex spacetime two-forms $u$ are not all independent. One condition arises from the requirement that
the holonomy of the supercovariant connection for all backgrounds is a subgroup of $SL(16, \bC)$. This
in particular gives
\bea
u_{MN,rq} B(\eta^r, \nu^q)=0~.
\la{conhol}
\eea
Taking this into account, the number of independent two forms $u$ for $N=2n$ supersymmetric backgrounds is equal  to the dimension
of $SL(16-n, \bC)\ltimes \oplus_n \bC^{16-n}$ as expected. In the cases we shall investigate below, the
basis $\eta^r$ is chosen in such a way that (\ref{conhol}) is automatically satisfied.

Apart from (\ref{conhol}), there are additional conditions
on the two-forms $u$. In particular those that arise from the Bianchi identities and field equations
of IIB supergravity described in the previous section. These can potentially further reduce the holonomy
of the supercovariant connection to a proper subgroup of $SL(16-n, \bC)\ltimes \oplus_n \bC^{16-n}$ .

In the special case for which $N=30$, and so $n=15$, that we are interested in, there is a unique (complex) normal direction
 $\nu$ to the
Killing spinors. The holonomy of the supercovariant connection is contained in $\bC^{15}$.
Taking into account the condition (\ref{conhol}), the supercovariant curvature is determined in terms
of 15 complex spacetime two-forms $u$, as expected. Furthermore, we shall show that all these 15 two-forms vanish subject to the
Bianchi identities and field equations of IIB supergravity. As a result ${\cal R}=0$ and $N=30$ IIB supergravity backgrounds
are locally maximally supersymmetric. There are three cases to consider depending on the orbit type of
the normal to the Killing spinors.

\newsection{$Spin(7)$-invariant normal}

The normal direction  can be chosen as $\nu = e_5 + e_{12345}$. A suitable basis such that
(\ref{conhol}) is automatically satisfied is
\bea
\eta^{\bar{\alpha} \bar{\beta}} = e_{\alpha \beta}
~,~~~
\eta^{\bar{\alpha}} &=& e_{\alpha5}~,
\notag\\
\eta^\alpha = {1 \over 6} \epsilon^{\alpha \beta_1 \beta_2 \beta_3}
e_{\beta_1 \beta_2 \beta_3 5} ~,~~~
\eta^+ &=& 1 - e_{1234}~,
\eea
where $\alpha, \beta=1,2,3,4$.
By considering the relation
\be
(T^2)_{P_1 P_2} =- {1 \over 16} u_r B(\eta^r, \Gamma_{P_1 P_2} \nu)~,
\ee
where  the form indices $MN$ have been suppressed in $(T^2)$ and in $u_r$,
we find the relations
\bea
(T^2)_{+-}=(T^2)_{- \mu}= (T^2)_{- \bar{\mu}}= 0~,~~~
(T^2)_{+ \mu} = -{1 \over 8} u_\mu~,~~~
(T^2)_{+ \bar{\mu}} &=& -{1 \over 8} u_{\bar{\mu}}~,
\notag\\
(T^2)_{\mu \nu} = -{1 \over 16} \epsilon_{\mu \nu}{}^{\bar{\beta}_1
\bar{\beta}_2} u_{\bar{\beta}_1 \bar{\beta}_2}~,~~~
(T^2)_{\mu \bar{\nu}} ={1 \over 8} u_+ \delta_{\mu \bar{\nu}}~,~~~
(T^2)_{\bar{\mu} \bar{\nu}} &=& {1 \over 8} u_{\bar{\mu}
\bar{\nu}}~.
\eea
Note that $u_{MN,r}$ are complex valued.
To proceed, observe that
\be
u_+ = 2 (T^2)_\alpha{}^\alpha
\ee
and hence, making use of the constraint
$(T^2_{MN})_{P_1P_2} = (T^2_{P_1P_2})_{MN}$,
we find that
\be
(T^2_{\alpha \bar{\beta}})_{\mu \bar{\nu}} =
{1 \over 16} (T^2{}_\rho{}^\rho)_\lambda{}^\lambda \delta_{\alpha
\bar{\beta}} \delta_{\mu \bar{\nu}}~.
\ee

Next note that (making use of $(T^2)_{- \mu}=0$)
\be
0 = (T^2{}_{N \bar{\beta}})_\mu{}^N = (T^2{}_{\sigma
\bar{\beta}})_\mu{}^\sigma +  (T^2{}_{\bar{\sigma}
\bar{\beta}})_\mu{}^{\bar{\sigma}}~.
\ee
However,

\be
 (T^2{}_{\bar{\sigma}
\bar{\beta}})_\mu{}^{\bar{\sigma}} = -{1 \over 16}
\epsilon_\mu{}^{\bar{\beta}_1
\bar{\beta}_2 \bar{\beta}_3} u_{\bar{\beta}_1
\bar{\beta}, \bar{\beta}_2 \bar{\beta}_3}
= -{1 \over 2} \epsilon_\mu{}^{\bar{\beta}_1
\bar{\beta}_2 \bar{\beta}_3} (T^2{}_{\bar{\beta}_1
\bar{\beta}})_{\bar{\beta}_2 \bar{\beta}_3}  =0
\ee
by the Bianchi identity. Hence, it follows that
$ (T^2{}_{\sigma
\bar{\beta}})_\mu{}^\sigma=0$, which implies that
$ (T^2{}_\rho{}^\rho)_\lambda{}^\lambda=0$. Hence
\be
(T^2{}_{\alpha \bar{\beta}})_{\mu \bar{\nu}} =0
\ee
so
\be
u_{\alpha \bar{\beta},+}=0~.
\ee
Similarly, we also have
\be
(T^2{}_{+ \alpha})_{\mu \bar{\nu}} = {1 \over 8} u_{+\alpha,+}
\delta_{\mu \bar{\nu}}
\ee
and hence $u_{+ \alpha,+} = 2 (T^2{}_{+ \alpha})_\lambda{}^\lambda$,
so
\be
(T^2{}_{+ \alpha})_{\mu \bar{\nu}} = {1 \over 4}
(T^2{}_{+ \alpha})_\lambda{}^\lambda  \delta_{\mu \bar{\nu}}~.
\ee

Next, note that
\be
0 = (T^2{}_{N+}){}_\mu{}^N
= (T^2{}_{\sigma +})_\mu{}^\sigma
+ (T^2{}_{\bar{\sigma} +})_\mu{}^{\bar{\sigma}}~,
\ee
where we have made use of $(T^2)_{+-}=0$.
However, $ (T^2{}_{\bar{\sigma} +})_\mu{}^{\bar{\sigma}}=0$
from the Bianchi identity, hence $ (T^2{}_{\sigma +})_\mu{}^\sigma=0$
also. This implies that $(T^2{}_{+\alpha})_\lambda{}^\lambda=0$, so $(T^2{}_{+\alpha})_{\mu \bar{\nu}}=0$.
Therefore $u_{+\alpha,+}=0$.
Also, $(T^2{}_{+\alpha})_{\mu \bar{\nu}}=0$ implies that $(T^2{}_{+ \bar{\alpha}})_{\mu \bar{\nu}}=0$
(as $T^2$ is real), hence it follows that $u_{+ \bar{\alpha},+}=0$.

The vanishing of $(T^2{}_{\mu \bar{\nu}})_{-\alpha}$, $(T^2{}_{\mu \bar{\nu}})_{- \bar{\alpha}}$,
and $(T^2{}_{\mu \bar{\nu}})_{+-}$ also implies that
$u_{- \alpha,+}=0$, $u_{- \bar{\alpha},+}=0$ and $u_{+-,+}=0$.
Next, consider
\be
(T^2{}_{\alpha \beta})_{\mu \bar{\nu}}= {1 \over 8}
u_{\alpha \beta,+} \delta_{\mu \bar{\nu}}~.
\ee
Contracting with $\epsilon^{\alpha \beta \mu}{}_{\bar{\lambda}}$
and using the Bianchi identity we find $u_{\alpha \beta,+}=0$,
so $(T^2{}_{\alpha \beta})_{\mu \bar{\nu}}=0$.
As $T^2$ is real, this implies that
$(T^2{}_{\bar{\alpha} \bar{\beta}})_{\mu \bar{\nu}}=0$, which then fixes
$u_{\bar{\alpha} \bar{\beta},+}=0$.
So all components of $u_+$ vanish.

Next, recall that $(T^2)_{+ \mu}=-{1 \over 8} u_\mu$.
Then the vanishing of $(T^2{}_{+ \mu})_{\alpha \bar{\beta}}$,
$(T^2{}_{+ \mu})_{- \alpha}$, $(T^2{}_{+ \mu})_{- \bar{\alpha}}$ and $(T^2{}_{+ \mu})_{-+}$
implies that
\be
u_{\alpha \bar{\beta}, \mu}=0, \qquad u_{- \alpha, \mu}=0, \qquad
u_{- \bar{\alpha}, \mu}=0, \qquad u_{-+, \mu}=0~.
\ee

Next note that
\be
(T^2{}_{\alpha \beta})_{+ \mu} = (T^2{}_{+ \mu})_{\alpha \beta}
= -{1 \over 2} \epsilon_{\alpha \beta}{}^{\bar{\rho} \bar{\sigma}}
(T^2{}_{+ \mu})_{\bar{\rho} \bar{\sigma}}~.
\ee
However, we also have $(T^2{}_{+ [\mu})_{\bar{\rho} \bar{\sigma}]}=0$.
Together with $(T^2)_{\mu \bar{\sigma}}=0$ this implies that
$(T^2{}_{+ \mu})_{\bar{\rho} \bar{\sigma}}=0$ and hence
$(T^2{}_{\alpha \beta})_{+ \mu}=0$ also. Hence $u_{\alpha \beta, \mu}=0$.
Furthermore, $(T^2{}_{\bar{\rho} \bar{\sigma}})_{+ \mu}=0$
implies that $u_{\bar{\alpha} \bar{\beta}, \mu}=0$ as well.

Next consider $(T^2)_{+ \bar{\mu}}=-{1 \over 8} u_{\bar{\mu}}$. The
vanishing of $(T^2{}_{+ \bar{\mu}})_{\alpha \bar{\beta}}$,
$(T^2{}_{+ \bar{\mu}})_{- \alpha}$, $(T^2{}_{+ \bar{\mu}})_{- \bar{\alpha}}$,
$(T^2{}_{+ \bar{\mu}})_{-+}$, $(T^2{}_{+ \bar{\mu}})_{\alpha \beta}$
and $(T^2{}_{\bar{\alpha} \bar{\beta}})_{+ \bar{\mu}}$
implies that
\be
u_{\alpha \bar{\beta}, \bar{\mu}}=0, \quad
u_{- \alpha, \bar{\mu}}=0, \quad u_{- \bar{\alpha}, \bar{\mu}}=0,
\quad  u_{-+, \bar{\mu}}=0, \quad u_{\alpha \beta, \bar{\mu}}=0, \quad
u_{\bar{\alpha} \bar{\beta}, \bar{\mu}}=0~.
\ee

Next consider the constraint $(T^2)_{\bar{\mu} \bar{\nu}}={1 \over 8}
u_{\bar{\mu} \bar{\nu}}$.
As
\be
(T^2{}_{\alpha {\bar{\beta}}})_{\bar{\mu} \bar{\nu}}=
(T^2{}_{\bar{\mu} \bar{\nu}})_{\alpha \bar{\beta}}=0~,
\ee
it follows that $u_{\alpha \bar{\beta}, \bar{\mu} \bar{\nu}}=0$.
Similarly, the vanishing of $(T^2{}_{\bar{\mu} \bar{\nu}})_{- \alpha}$,
$(T^2{}_{\bar{\mu} \bar{\nu}})_{- \bar{\alpha}}$ ,
$(T^2{}_{\bar{\mu} \bar{\nu}})_{+-}$,
$(T^2{}_{\bar{\mu} \bar{\nu}})_{+ \alpha}$
and $(T^2{}_{\bar{\mu} \bar{\nu}})_{+ \bar{\alpha}}$
implies that
\be
u_{- \alpha, \bar{\mu} \bar{\nu}}=0, \quad
u_{- \bar{\alpha}, \bar{\mu} \bar{\nu}}=0, \quad
u_{+-, \bar{\mu} \bar{\nu}}=0, \quad
u_{+ \alpha, \bar{\mu} \bar{\nu}}=0, \quad
u_{+ \bar{\alpha}, \bar{\mu} \bar{\nu}}=0~.
\ee

Next consider the Bianchi identity
\be
(T^2{}_{\alpha [\beta})_{\bar{\mu} \bar{\nu}]}=0~.
\ee
As $u_+=0$, it follows that
$(T^2{}_{\alpha \bar{\nu}})_{\beta \bar{\nu}}=0$,
and hence $(T^2{}_{\alpha \beta})_{\bar{\mu} \bar{\nu}}=0$.
Therefore $u_{\alpha \beta, \bar{\mu} \bar{\nu}}=0$.
Also
\be
(T^2_{\bar{\alpha} \bar{\beta}})_{\bar{\mu} \bar{\nu}}
= -{1 \over 2} \epsilon_{\bar{\mu} \bar{\nu}}{}^{\lambda_1
\lambda_2} (T^2{}_{\bar{\alpha} \bar{\beta}})_{\lambda_1 \lambda_2}
=0~,
\ee
so $u_{\bar{\alpha} \bar{\beta}, \bar{\mu} \bar{\nu}}=0$.
Hence all components of $u_{\bar{\mu} \bar{\nu}}$ vanish.

To summarize, these constraints fix all components of $u_r$ to vanish,
with the exception of $u_{+A,B}$ where $A, B$ are $su(4)$
indices. As
\be
\label{spin7con1}
(T^2{}_{+A})_{+B} = -{1 \over 8} u_{+A,B}~,
\ee
it follows that $u_{+A,B}$ is symmetric in $A,B$.

Next consider the 4-forms. It turns out that
all components of $T^4$ are forced to vanish
by the above constraints with the exception of

\bea
(T^4)_{+\mu \nu \rho} &=& -{1 \over 4} u_{\bar{\alpha}}
\epsilon^{\bar{\alpha}}{}_{\mu \nu \rho}~,~~~
(T^4)_{+ \mu \nu \bar{\rho}} = {1 \over 8} u_\mu
\delta_{\nu \bar{\rho}} - {1 \over 8} u_\nu \delta_{\mu \bar{\rho}}~,
\notag\\
(T^4)_{+ \mu \bar{\nu} \bar{\rho}} &=&
-{1 \over 8} \delta_{\mu \bar{\nu}} u_{\bar{\rho}}
+{1 \over 8} \delta_{\mu \bar{\rho}} u_{\bar{\nu}}
~,~~~
(T^4)_{+ \bar{\mu} \bar{\nu} \bar{\rho}} = -{1 \over 4}
u_\alpha \epsilon^\alpha{}_{\bar{\mu} \bar{\nu} \bar{\rho}}~.
\eea
Using ({\ref{spin7con1}}), this implies that

\bea
\label{spin7con2}
(T^4)_{+\mu \nu \rho} &=& 2 (T^2)_{+ \bar{\alpha}}
\epsilon^{\bar{\alpha}}{}_{\mu \nu \rho}~,~~~
(T^4)_{+ \mu \nu \bar{\rho}} = (T^2)_{+ \nu} \delta_{\mu \bar{\rho}}
-(T^2)_{+ \mu} \delta_{\nu \bar{\rho}}~,
\notag\\
(T^4)_{+ \bar{\mu} \bar{\nu} \rho}
&=& (T^2)_{+ \bar{\nu}} \delta_{\bar{\mu} \rho}
-(T^2)_{+ \bar{\mu}} \delta_{\bar{\nu} \rho}~,~~~
(T^4)_{+ \bar{\mu} \bar{\nu} \bar{\rho}} = 2 (T^2)_{+ \alpha}
\epsilon^\alpha{}_{\bar{\mu} \bar{\nu} \bar{\rho}}~.
\eea
This implies that $T^4$ is entirely real, so that $F$ is covariantly constant.
Furthermore, $(T^4{}_{+A_1})_{+A_2 A_3 A_4}$
is totally antisymmetric in $A_1, A_2, A_3, A_4$.
Recall that $(T^4{}_{M[P_1})_{P_2 P_3 P_4 P_5]}$
is self-dual in the five anti-symmetrized indices.
Hence $(T^4{}_{+\alpha_1})_{+ \alpha_2 \alpha_3 \alpha_4}$
must vanish. Then ({\ref{spin7con2}}) implies that
$(T^2{}_{+\alpha})_{+ \bar{\beta}}=0$.

Also consider
\be
(T^4{}_{+ \alpha}){}_{+ \mu  \nu \bar{\rho}}
= \delta_{\mu \bar{\rho}} (T^2{}_{+ \alpha})_{+ \nu}
- \delta_{\nu \rho} (T^2{}_{+ \alpha})_{+ \mu}~.
\ee
Contracting this identity gives
\be
(T^4{}_{+ \alpha})_{+ \mu \lambda}{}^\lambda = -3 (T^2{}_{+ \alpha})_{+ \mu}~.
\ee
However, the self-duality condition implies that
$ (T^4{}_{+ \alpha})_{+ \mu \lambda}{}^\lambda=0$, and hence
$(T^2{}_{+ \alpha})_{+ \beta}=0$ also. Therefore, all components
of $T^2$ and $T^4$ are constrained to vanish.


 \newsection{$SU(4)\ltimes\bR^8$-invariant normal}

The normal spinor direction is taken to be

\be
\nu = (n-\ell +im) e_5 +(n+\ell+im)e_{12345}~,
\ee
and a basis in the space of Killing spinors such that (\ref{conhol}) is satisfied is
\bea
\eta^{\bar{\alpha} \bar{\beta}} &=& e_{\alpha \beta}~,~~~
\eta^{\bar{\alpha}} =e_{\alpha 5}~,
\notag\\
\eta^\alpha &=& {1 \over 6} \epsilon^{\alpha \beta_1 \beta_2 \beta_3}
e_{\beta_1 \beta_2 \beta_3 5} ~,~~~
\eta^+ = (n-\ell+im)1 - (n+\ell+im)e_{1234}~.
\eea
$T^2$ is constrained by
\bea
\label{su4t2}
(T^2)_{+-} &=&(T^2)_{- \mu}=(T^2)_{- \bar{\mu}} = 0~,
\notag\\
(T^2)_{+ \mu} &=& -{1 \over 8} (n-\ell +im) u_\mu~,~~~
(T^2)_{+ \bar{\mu}} = -{1 \over 8} (n+\ell+im) u_{\bar{\mu}}~,
\notag\\
(T^2)_{\mu \nu} &=& -{1 \over 16} (n-\ell +im) \epsilon_{\mu \nu}{}^{\bar{\beta}_1
\bar{\beta}_2} u_{\bar{\beta}_1 \bar{\beta}_2}~,~~~
(T^2)_{\mu \bar{\nu}} = {1 \over 8} \big( (n+im)^2-\ell^2 \big)  u_+ \delta_{\mu \bar{\nu}}~,
\notag\\
(T^2)_{\bar{\mu} \bar{\nu}} &=& {1 \over 8} (n+\ell +im) u_{\bar{\mu}
\bar{\nu}}~.
\eea
The analysis proceeds depending on whether or not $ (n+im)^2-\ell^2$ vanishes. There are three cases
but two of them are related by a $Spin(9,1)$ transformation. So there are two
independent cases to consider.

\subsection{Generic solutions ($ (n+im)^2-\ell^2 \neq 0$)}

In this case there are no restrictions on the spacetime functions $n,m$ and $\ell$.
It is then straightforward to see,
using the same reasoning as in the $Spin(7) \ltimes\bR^8$ analysis, that all components of
$u_r$ vanish except for $u_{+A,B}$, where $A=(\a, \bar\a)$, $B=(\beta, \bar\beta)$, and

\bea
(T^2{}_{+ \alpha})_{+ \beta} &=& -{1 \over 8} (n-\ell+im) u_{+\alpha,\beta}~,~~~
(T^2{}_{+ \alpha})_{+ \bar{\beta}} = -{1 \over 8} (n+\ell+im) u_{+\alpha,\bar{\beta}}~,
\notag\\
(T^2{}_{+ \bar{\alpha}})_{+ \beta} &=& -{1 \over 8} (n-\ell+im) u_{+\bar{\alpha},\beta}~,~~~
(T^2{}_{+ \bar{\alpha}})_{+ \bar{\beta}} = -{1 \over 8} (n+\ell+im) u_{+\bar{\alpha},\bar{\beta}}~.
\eea
Similarly,  it turns out that
all components of $T^4$ are forced to vanish
by the above constraints with the exception of
\bea
(T^4)_{+\mu \nu \rho} &=& -{1 \over 4}(n-\ell+im) u_{\bar{\alpha}}
\epsilon^{\bar{\alpha}}{}_{\mu \nu \rho}~,
\notag\\
(T^4)_{+ \mu \nu \bar{\rho}} &=& {1 \over 8} (n-\ell+im) \big( u_\mu
\delta_{\nu \bar{\rho}} - u_\nu \delta_{\mu \bar{\rho}} \big)~,
\notag\\
(T^4)_{+ \mu \bar{\nu} \bar{\rho}} &=&
{1 \over 8} (n+\ell+im) \big( \delta_{\mu \bar{\rho}} u_{\bar{\nu}}
-\delta_{\mu \bar{\nu}} u_{\bar{\rho}} \big)~,
\notag\\
(T^4)_{+ \bar{\mu} \bar{\nu} \bar{\rho}} &=& -{1 \over 4}(n+\ell+im)
u_\alpha \epsilon^\alpha{}_{\bar{\mu} \bar{\nu} \bar{\rho}}~.
\eea

As $(T^4{}_{+A_1})_{+A_2 A_3 A_4}$ is totally antisymmetric in $A_i$,
self-duality implies that $(T^4{}_{+\alpha})_{+\beta \rho \sigma}=0$,
and hence $u_{+\alpha, \bar{\beta}}=0$. Therefore $(T^2{}_{+\alpha})_{+ \bar{\beta}}=0$,
and hence $(T^2{}_{+ \bar{\alpha}})_{+ \beta}=0$ also implies $u_{+\bar{\alpha},\beta}=0$.

Furthermore, we also have
\be
(T^4{}_{+\mu})_{+\alpha \beta}{}^\beta = {3 \over 8} (n-\ell+im) u_{+\mu, \alpha}~.
\ee
As the left-hand side of this expression must vanish by self-duality, we find
$u_{+\alpha,\beta}=0$. Hence $(T^2{}_{+\alpha})_{+\beta}=0$, and so
$(T^2{}_{+\bar{\alpha}})_{+\bar{\beta}}=0$ also implies that $u_{+\bar{\alpha},\bar{\beta}}=0$.
Therefore all components of the $u_r$ vanish, so all components of $T^2$ and $T^4$ are constrained
to vanish as well.

\subsection{Pure spinor solution ($ (n+im)^2-\ell^2 = 0$)}

There are two pure spinor cases that one can consider depending on whether
 $m=0$, $n=\ell\not=0$ or  $m=0$, $n=-\ell\not=0$.  The normal directions are either $\nu=e_{1234}$ or $\nu=1$, respectively.
 However, these two normals are related by a $Spin(9,1)$ transformation. So it suffices to consider one of the two cases as
 the other will follow by virtue of the $Spin(9,1)$ gauge symmetry of the  Killing spinor equations.
So let us investigate the case
 $m=0$, $n=\ell$. Then ({\ref{su4t2}}) implies that $(T^2)_{+ \alpha}=0$.
Therefore, $(T^2)_{+ \bar{\alpha}}=0$, so $u_{\bar{\alpha}}=0$. Furthermore,
$(T^2)_{\alpha \beta}=0$, so $(T^2)_{\bar{\alpha} \bar{\beta}}=0$ also, and therefore
$u_{\bar{\alpha} \bar{\beta}}=0$. These constraints are sufficient to fix $T^2=0$, however
$u_+$ and $u_\alpha$ are not fixed by constraints involving $T^2$.

It is straightforward to see that the only non-vanishing components of $T^4$ are given by
\bea
(T^4)_{+ \bar{\alpha} \bar{\beta} \bar{\lambda}} &=& {n \over 2}
\epsilon_{ \bar{\alpha} \bar{\beta} \bar{\lambda}}{}^\rho u_\rho~,~~~
(T^4)_{\bar{\alpha} \bar{\beta} \bar{\lambda} \bar{\sigma}} = -n^2 u_+
\epsilon_{\bar{\alpha} \bar{\beta} \bar{\lambda} \bar{\sigma}}~.
\eea

To proceed, note that the self-duality constraint fixes $(T^4{}_{+ \bar{\sigma}})_{+ \bar{\alpha} \bar{\beta} \bar{\lambda}}=0$,
so $u_{+ \bar{\beta},\alpha}=0$.
Also, $(T^4{}_{+ \sigma})_{+ \bar{\alpha} \bar{\beta} \bar{\lambda}}=-
(T^4{}_{+ \bar{\alpha}})_{+\sigma \bar{\beta} \bar{\lambda}}=0$, so $u_{+ \beta, \alpha}=0$.
Furthermore $(T^4{}_{[\mu \nu})_{\bar{\alpha} \bar{\beta} \bar{\lambda} \bar{\sigma}]}=0$
which implies $(T^4{}_{\mu \nu})_{\bar{\alpha} \bar{\beta} \bar{\lambda} \bar{\sigma}}=0$
and hence $u_{\mu \nu,+}=0$.
Also, $(T^4{}_{[- \nu})_{\bar{\alpha} \bar{\beta} \bar{\lambda} \bar{\sigma}]}=0$
implies $(T^4{}_{- \nu})_{\bar{\alpha} \bar{\beta} \bar{\lambda} \bar{\sigma}}=0$,
so $u_{-\alpha,+}=0$.

Next, consider the following relation implied by self-duality:
\be
(T^4{}_{+ [\nu})_{\bar{\alpha} \bar{\beta} \bar{\lambda} \bar{\sigma}]}
= -{1 \over 6} \epsilon_{\bar{\alpha} \bar{\beta} \bar{\lambda} \bar{\sigma}}
\epsilon_\nu{}^{\bar{\lambda}_1 \bar{\lambda}_2 \bar{\lambda}_3}
(T^4{}_{+[-})_{+ \bar{\lambda}_1 \bar{\lambda}_2 \bar{\lambda}_3]}~.
\ee
This implies that
\be
n u_{+ \alpha,+} = -{1 \over 2} u_{+-,\alpha}~.
\ee
However, $(T^4{}_{+-})_{+ \bar{\lambda}_1 \bar{\lambda}_2 \bar{\lambda}_3}
= -(T^4{}_{+ \bar{\lambda}_1})_{+- \bar{\lambda}_2 \bar{\lambda}_3} =0$,
which implies that $u_{+-,\alpha}=0$, so $u_{+ \alpha,+}=0$ as well.
Also, $(T^4{}_{[- \rho})_{+ \bar{\alpha} \bar{\beta} \bar{\lambda}]}=0$,
which implies $(T^4{}_{- \rho})_{+ \bar{\alpha} \bar{\beta} \bar{\lambda}}=0$
and so $u_{- \alpha, \beta}=0$.

Also note that $(T^4{}_{- (\bar{\alpha}})_{\bar{\beta}) \bar{\rho} \bar{\sigma}
\bar{\lambda}}= -(T^4{}_{\bar{\rho} (\bar{\alpha}})_{\bar{\beta}) - \bar{\sigma} \bar{\lambda}}=0$,
so
\be
u_{- \bar{\alpha},+} \epsilon_{\bar{\beta} \bar{\rho} \bar{\sigma} \bar{\lambda}}
+ u_{- \bar{\beta},+} \epsilon_{\bar{\alpha} \bar{\rho} \bar{\sigma} \bar{\lambda}}=0~.
\ee
Contracting this expression with $\epsilon^{\bar{\beta} \bar{\rho} \bar{\sigma} \bar{\lambda}}$
yields $u_{- \bar{\alpha},+}=0$.

Next consider $(T^4{}_{- (+})_{\bar{\alpha}) \bar{\beta} \bar{\lambda} \bar{\sigma}}
= - (T^4{}_{\bar{\beta} (+})_{\bar{\alpha}) - \bar{\lambda} \bar{\sigma}}=0$.
This implies that
\be
n^2 u_{-+,+} \epsilon_{\bar{\alpha} \bar{\beta} \bar{\lambda} \bar{\sigma}}
-{n \over 2} u_{- \bar{\alpha}, \rho} \epsilon_{\bar{\beta} \bar{\lambda} \bar{\sigma}}{}^\rho =0
\ee
and on contracting with $\epsilon^{\bar{\beta} \bar{\lambda} \bar{\sigma}}{}_\mu$, we find
\be
\label{su4aux1}
u_{- \bar{\alpha},\mu} = -2n \delta_{\bar{\alpha} \mu} u_{-+,+}~.
\ee
However, self-duality implies that $(T^4{}_{-[+})_{\bar{\alpha} \bar{\beta} \bar{\lambda}
\bar{\sigma}]}=0$, which when combined with ({\ref{su4aux1}}) is sufficient to constrain
$u_{-+,+}=0$ and hence $u_{- \bar{\alpha},\mu}=0$ as well.

Next, note that $(T^4{}_{\mu (\bar{\nu}})_{\bar{\alpha}) \bar{\beta} \bar{\lambda} \bar{\rho}}
= - (T^4{}_{\bar{\beta} (\bar{\nu}})_{\bar{\alpha}) \mu \bar{\lambda} \bar{\rho}}=0$,
hence
\be
u_{\mu \bar{\nu},+} \epsilon_{\bar{\alpha} \bar{\beta} \bar{\lambda} \bar{\rho}}
+ u_{\mu \bar{\alpha},+} \epsilon_{\bar{\nu} \bar{\beta} \bar{\lambda} \bar{\rho}} =0~.
\ee
On contracting this identity with $\epsilon^{\bar{\alpha} \bar{\beta} \bar{\lambda} \bar{\rho}}$
we find $u_{\mu \bar{\nu},+}=0$.

The constraint $(T^4{}_{+ (\bar{\mu}})_{\bar{\alpha}) \bar{\beta} \bar{\lambda} \bar{\sigma}}
= - (T^4{}_{\bar{\beta} ({\bar{\mu}}})_{\bar{\alpha}) + \bar{\lambda} \bar{\sigma}}$
implies, on contracting with $\epsilon^{\bar{\alpha} \bar{\beta} \bar{\lambda}
\bar{\sigma}}$, that
\be
\label{su4aux2}
6n u_{+ \bar{\mu},+} = - \delta^{\rho \bar{\beta}} u_{\bar{\beta} \bar{\mu}, \rho}
\ee
and furthermore the self-duality constraint $(T^4{}_{\bar{\mu} [+})_{\bar{\alpha} \bar{\beta}
\bar{\lambda} \bar{\sigma}]}=0$ implies, on contracting with
 $\epsilon^{\bar{\alpha} \bar{\beta} \bar{\lambda}
\bar{\sigma}}$, that
\be
24 n^2 u_{+ \bar{\mu},+} -12n \delta^{\rho \bar{\beta}} u_{\bar{\beta} \bar{\mu}, \rho}=0~.
\ee
This constraint, together with ({\ref{su4aux2}}) implies that
$u_{+, \bar{\mu},+}=0$ and $\delta^{\rho \bar{\beta}} u_{\bar{\beta} \bar{\mu}, \rho}=0$.
Next note that $(T^4{}_{\bar{\mu} (\bar{\nu}})_{\bar{\alpha}) \bar{\beta} \bar{\rho} \bar{\sigma}}
= - (T^4{}_{\bar{\beta} (\bar{\nu}})_{\bar{\alpha}) \bar{\mu} \bar{\rho} \bar{\sigma}}$.
Contracting this constraint with $\epsilon^{\bar{\alpha} \bar{\beta} \bar{\rho} \bar{\sigma}}$
gives $u_{\bar{\mu} \bar{\nu},+}=0$.

Combining all of these constraints fixes all components of $u_+$ to vanish. To fix the
remaining components of $u_{\alpha}$, note that $(T^4{}_{\bar{\mu} (\bar{\nu}})_{+) \bar{\alpha}
\bar{\beta} \bar{\lambda}} = -(T^4{}_{\bar{\alpha} (\bar{\nu}})_{+) \bar{\mu} \bar{\beta} \bar{\lambda}}$
implies that
\be
\epsilon_{\bar{\alpha} \bar{\beta} \bar{\lambda}}{}^\rho u_{\bar{\mu} \bar{\nu}, \rho}
= - \epsilon_{\bar{\mu} \bar{\beta} \bar{\lambda}}{}^\rho u_{\bar{\alpha} \bar{\nu}, \rho}
\ee
and on contracting this expression with $\epsilon^{\bar{\alpha} \bar{\beta} \bar{\lambda}}{}_\sigma$
and using the constraint  $\delta^{\rho \bar{\beta}} u_{\bar{\beta} \bar{\mu}, \rho}=0$
which we have already obtained, we find $u_{\bar{\mu} \bar{\nu}, \sigma}=0$.

Next, note that the constraint $(T^4{}_{\mu (\bar{\nu}})_{+) \bar{\alpha} \bar{\beta} \bar{\lambda}}
= -(T^4{}_{\bar{\alpha} (\bar{\nu}})_{+) \mu \bar{\beta} \bar{\lambda}} =0$
together with $u_+=0$ implies that $(T^4{}_{\mu \bar{\nu}})_{+ \bar{\alpha} \bar{\beta} \bar{\lambda}}=0$,
so $u_{\mu \bar{\nu}, \rho}=0$.
Finally, $(T^4{}_{\mu (\nu})_{+) \bar{\alpha} \bar{\beta} \bar{\lambda}}
= - (T^4{}_{\bar{\alpha} (\nu})_{+) \mu \bar{\beta} \bar{\lambda}} =0$
together with $u_+=0$ imply that $(T^4{}_{\mu \nu})_{+ \bar{\alpha} \bar{\beta} \bar{\lambda}}=0$,
so $u_{\mu \nu, \rho}=0$.

These constraints are then sufficient to fix $u_\alpha=0$, and hence all components of $u_r$
vanish, as do $T^2$ and $T^4$.

\newsection{$G_2$-invariant normal}

The normal spinor can be  chosen as
\be
\nu = n (e_5+e_{12345})+im (e_1+e_{234})~.
\ee
By using a gauge transformation of the form $e^{f \Gamma_{+-}}$
for real $f$, we can without loss of generality set $m=\pm n$,
and so we take the normal spinor direction as
\be
\nu = e_5+e_{12345} \pm i (e_1+e_{234})~.
\ee
A basis of   spinors compatible with (\ref{conhol}) is
\bea
\eta^- &=& e_{15}+e_{2345} \mp i (1+e_{1234})~,~~~
\eta^+ = 1-e_{1234}~,
\notag\\
\eta^1 &=& e_{15}-e_{2345}~,~~~
\eta^{1 \bar{p}} = e_{1p}~,~~~\eta^{1p} ={1 \over 2} \epsilon_{pqr} e_{qr}~,
\notag\\
\eta^{\bar{p}} &=&e_{p5}~,~~~
\eta^p = {1 \over 2} \epsilon_{pqr} e_{qr} \wedge e_{15}~,
\eea
where $p,q,r=1,2,3$.
We then find the following constraints on $T^2$:

\beann
(T^2)_{+-} &=& \pm {i \over 4} u_-~,~~~
(T^2)_{+1} = -{1 \over 8}(u_- - u_1)~,~~~
(T^2)_{+ \bar{1}} = -{1 \over 8} (u_-+u_1)~,
\notag\\
(T^2)_{+p} &=& {1 \over 8} u_p~,~~~
(T^2)_{+ \bar{p}} = -{1 \over 8} u_{\bar{p}}~,~~~
\notag\\
(T^2)_{-1} &=& -{1 \over 8}(-u_- \mp i u_+)~,~~~
(T^2)_{- \bar{1}} = -{1 \over 8}(-u_- \pm i u_+)~,~~~
(T^2)_{-p} = \pm {i \over 8} u_{1p}~,
\notag\\
(T^2)_{- \bar{p}} &=& \mp {i \over 8} u_{1 \bar{p}}~,
\eeann

\bea
\label{g2con1}
(T^2)_{1 \bar{1}} &=& -{1 \over 8}(\pm i u_1 - u_+)~,~~~
(T^2)_{1p} =-{1 \over 8} u_{1p}~,~~~
(T^2)_{1 \bar{p}} =\mp {i \over 8}u_{\bar{p}}~,
\notag\\
(T^2)_{\bar{1} p} &=& \pm {i \over 8} u_p~,~~~
(T^2)_{\bar{1} \bar{p}} = {1 \over 8} u_{1 \bar{p}}~,
\notag\\
(T^2)_{pq} &=& -{1 \over 8} \epsilon_{pq}{}^{\bar{r}}(
u_{1 \bar{r}} \pm i u_{\bar{r}})~,~~~
(T^2)_{p \bar{q}} = -{1 \over 8} \delta_{p \bar{q}}(-u_+ \mp i u_1)~,
\notag\\
(T^2)_{\bar{p} \bar{q}} &=& -{1 \over 8} \epsilon_{\bar{p} \bar{q}}{}^r
(- u_{1r} \mp i u_r)~.
\eea
These constraints imply that
\bea
\label{g2con2}
u_- &=& \mp 4i (T^2)_{+-}~,~~~
u_1 = -4((T^2)_{+ \bar{1}}-(T^2)_{+1})~,~~~
u_p = 8 (T^2)_{+p}~,
\notag\\
u_{\bar{p}} &=& -8 (T^2)_{+ \bar{p}}~,~~~
u_+ = \pm 4i ((T^2)_{- \bar{1}}-(T^2)_{-1})~,~~~
u_{1p} = -8 (T^2)_{1p}~,
\notag\\
u_{1 \bar{p}} &=& 8 (T^2)_{\bar{1} \bar{p}}~.
\eea
Substituting ({\ref{g2con2}}) back into ({\ref{g2con1}}) gives the
constraints
\beann
(T^2)_{+1} + (T^2)_{+ \bar{1}} &=& \pm i (T^2)_{+-}~,~~~
(T^2)_{-1} + (T^2)_{- \bar{1}} = \mp i (T^2)_{+-}~,
\notag\\
(T^2)_{-p} &=& \mp i (T^2)_{1p}~,~~~
(T^2)_{- \bar{p}} = \mp i (T^2)_{\bar{1} \bar{p}}~,
\notag\\
(T^2)_{1 \bar{1}} &=& \pm {i \over 2}\big( (T^2)_{+ \bar{1}}
-(T^2)_{+1} + (T^2)_{- \bar{1}} - (T^2)_{-1} \big)~,
\eeann
\bea
(T^2)_{1 \bar{p}} &=& \pm i (T^2)_{+ \bar{p}}~,~~~
(T^2)_{\bar{1} p} = \pm i (T^2)_{+p}~,
\notag\\
(T^2)_{pq} &=& \epsilon_{pq}{}^{\bar{r}} (-(T^2)_{\bar{1} \bar{r}}
\pm i (T^2)_{+ \bar{r}})~,
\notag\\
(T^2)_{p \bar{q}} &=& \pm {i \over 2} \delta_{p \bar{q}}
\big( (T^2)_{- \bar{1}} - (T^2)_{-1} - (T^2)_{+ \bar{1}}
+ (T^2)_{+1} \big)~,
\notag\\
(T^2)_{\bar{p} \bar{q}} &=& \epsilon_{\bar{p} \bar{q}}{}^r
(-(T^2)_{1r} \pm i (T^2)_{+r})~.
\eea
These constraints can be rewritten in terms of irreducible $G_2$ representations\footnote{This can be seen as
a consistency check of the calculation.} as
\bea
(T^2)_{+\underline 1}&=&\pm \tfrac{i}{\sqrt{2}}(T^2)_{+-}~,~~~(T^2)_{-\underline 1}=\mp\tfrac{i}{\sqrt{2}}(T^2)_{+-}~,\notag\\
(T^2)_{\underline{1}i}&=&\pm \tfrac{i}{\sqrt{2}}((T^2)_{+i}+(T^2)_{-i})~,\notag\\
(\Pi_{\bf 7} T^2)_{i}&:=&\varphi_{i}{}^{jk}(T^2)_{jk}=\pm 3\sqrt{2}i((T^2)_{+i}-(T^2)_{-i})~,\notag\\
(\Pi_{\bf 14}T^2)_{ij}&:=&{2\over3} ({1\over4} \star\varphi_{ij}{}^{kl} (T^2)_{kl}
\,
+(T^2)_{ij})=0~,
\eea
where the underlined 1 denotes a real index. By taking the complex conjugate of these expressions, and using the fact that $T^2_{MN}$ is real, one immediately finds that all components of $T^2_{MN}$ are put to zero. This implies, through (\ref{g2con2}), that all components of $u_r$ vanish.

Note that throughout this reasoning, in contrast to the analysis of
the $Spin(7) \ltimes \bR^8$ and $SU(4) \ltimes \bR^8$ cases,
we have not made use of the algebraic constraints
on $T^2$ given in ({\ref{physcond}}) and ({\ref{feqcond}});
only the fact that $T^2$ is real has been used.

To summarize, we have shown that all components of the $u_r$ vanish, so all components of $T^2$ and $T^4$ also vanish. This yields ${\cal R}=0$ in this case as well. We therefore conclude that for the
$N>28$ IIB backgrounds ${\cal R}=0$ and they are thus locally isometric to maximally supersymmetric backgrounds.

\newsection{Discrete quotients}

We have shown that all $N>28$ supersymmetric IIB backgrounds are locally maximally supersymmetric. So
it remains to exclude the possibility that $28<N<32$ backgrounds can be constructed by discrete quotients
of maximally supersymmetric ones. The maximally supersymmetric backgrounds of IIB supergravity have been classified \cite{gpjose}.
It has been found that they are locally isometric to
 Minkowski space $\bR^{9,1}$, $AdS_5\times S^5$ \cite{schwarz} and the maximally supersymmetric plane wave \cite{mspw}.
 Considering the simply connected maximally supersymmetric backgrounds, which we collectively denote as $\tilde M$,
 one chooses a discrete subgroup $D$ of their symmetry group $S$, and constructs new solutions by taking the quotient of $\tilde M$
 with $D$, $\tilde M/D$. Such backgrounds are solutions of the field equations and depending on the choice of $D$ typically preserve
 less supersymmetry than $\tilde M$. So the task is to find whether there are subgroups $D$ such that $\tilde M/D$ preserves
 $28<N<32$ supersymmetries. The linearity of the Killing spinor equations of IIB supergravity for backgrounds with $P=G=0$ over the
 complex numbers excludes the possibility of $\tilde M/D$ preserving an odd number of supersymmetries. So to prove
 that there are no new supersymmetric backgrounds with $N>28$, we have to show that there are no $N=30$ quotients of
 maximally supersymmetric backgrounds.

 The task of proving that there are no subgroups $D\subset S$ of the symmetry group of
 simply connected maximally supersymmetric IIB backgrounds  $\tilde M$ for
 which $\tilde M/D$ preserves 30 supersymmetries is simplified in two ways. First it has been shown in \cite{N=31-M2} that,
 without loss of generality, one
can consider only  cyclic subgroups  $D$  as the remaining possibilities can be reduced to  this case.
In addition, it suffices to take the generator $\a$ of the cyclic group, $D=<\a>$,  to lie in the image
of the exponential map of $S$. Therefore $\a=e^X$, where $X$ is an element of the Lie algebra of $S$.
 Since $D$ is specified up to a conjugation in $S$, it suffices to consider the normal forms of $X$ up to
the action of the adjoint map of $S$. This is a straightforward task for compact groups but for non-compact
ones, like $S$, there are several possibilities as has been emphasized in \cite{simona}.

One continues the  computation  by considering the lift $\hat\a$ of the
generator $\a$ to the spin bundle
and by computing the number of invariant Killing spinors under the action of $\hat\a$. The number of invariant Killing spinors
is the number of supersymmetries preserved by $\tilde M/D$.

One difference that arises in the IIB case, in comparison with the  cases investigated in \cite{simona, simonb}, is
that the group action should be lifted to a $Spin_c(9,1)=Spin(9,1)\times_{\bZ_2} U(1)$  rather than
a $Spin(9,1)$ bundle. This is equivalent to allowing an additional phase in the lift $\hat\a$ of the generator $\a$ of
$D$
along the $U(1)$ direction. This additional phase is similar to that which appears in the context
of supersymmetric backgrounds in three-dimensional supergravities as the holonomy
of a flat $U(1)$ connection \cite{gptownhowe}.
It is known that the inclusion of the $U(1)$  phase changes the number of supersymmetries preserved by a background. Such
backgrounds are the stringy cosmic strings \cite{vafa},  the D7-branes \cite{perry} and the conical purely gravitational
domain walls of \cite{gran}.

\subsection{Discrete quotients of $\bR^{9,1}$}

Let us begin with the flat space case. The translations do not reduce supersymmetry so they are not appropriate
for the construction of $N<32$ backgrounds. On the other hand discrete quotients with elements
of the isometry group $SO(9,1)$ of $\bR^{9,1}$ do not preserve all supersymmetry. So consider
the generator $\alpha=\exp X$, $X\in \mathfrak{so}(9,1)$, of the cyclic group. Then up to a conjugation,
one has that either
\bea
X=-\theta_0 e^0\wedge e^5+\theta_1 e^1\wedge e^6+\theta_2 e^2\wedge e^7+\theta_3 e^3\wedge e^8+\theta_4 e^4\wedge e^9~,
\eea
or
\bea
X= -(e^0-e^5)\wedge e^9+\theta_1 e^1\wedge e^6+\theta_2 e^2\wedge e^7+\theta_3 e^3\wedge e^8~.
\la{nulla}
\eea
In the former case, $\alpha$ lifts to the element
\bea
\hat\alpha= \exp\big({1\over2} (\theta_0\Gamma_{05}+\theta_1 \Gamma_{16}+\theta_2 \Gamma_{27}
+\theta_3 \Gamma_{38}+ \theta_4 \Gamma_{49})+i\psi\big)
\eea
of $Spin_c(9,1)$, where $\psi$ is the angle along the $U(1)$ direction. Since $\Gamma_{05}, \Gamma_{16}, \Gamma_{27}, \Gamma_{38}$ and $\Gamma_{49}$ are commuting
with $-(\Gamma_{05})^2= (\Gamma_{16})^2=(\Gamma_{27})^2= (\Gamma_{38})^2=(\Gamma_{49})^2=-1_{16\times16}$, the  Weyl
representation decomposes in subspaces which are the eigenspaces of the above matrices, i.e.
\bea
\Delta_{\bf 16}=\oplus_{\sigma_0,\dots , \sigma_4} W_{\sigma_0\dots \sigma_4}~,
\eea
where $\sigma_0,\dots, \sigma_4$ are signs restricted by the chirality condition to satisfy
$\sigma_0 \sigma_1\sigma_2\sigma_3\sigma_4=1$.
 Therefore
acting on the subspace $ W_{\sigma_0\dots \sigma_4}$, one has
\bea
\hat\alpha(\sigma_0,\dots, \sigma_4)=\exp\big({1\over2} (\sigma_0\theta_0+i \sigma_1\theta_1 +i \sigma_2\theta_2
+i \sigma_3\theta_3 +i \sigma_4 \theta_4 )+i\psi\big)~.
\eea
 Now to find  the supersymmetry  preserved by a discrete
quotient constructed from $\alpha$, one has to determined the spinors which are left invariant under the action
of $\hat\alpha$.
This in particular implies
that there must be angles or boosts such that
\bea
\exp\big({1\over2} (\sigma_0\theta_0+i \sigma_1\theta_1 +i \sigma_2\theta_2
+i \sigma_3\theta_3 +i \sigma_4 \theta_4 )+i\psi\big)=1~,
\eea
for some choice of signs $\sigma$.
Taking the complex conjugate, we conclude that
\bea
\theta_0=0~.
\eea
Moreover, since we require at least 30 supersymmetries to be preserved, there are $\sigma_0,\sigma_1,\dots, \sigma_4$ such
that if $\hat\alpha(\sigma_0,\sigma_1,\dots, \sigma_4)=1$, then $\hat\alpha(\sigma_0,\bar\sigma_1,\dots, \bar\sigma_4)=1$ for
 $\bar \sigma=-\sigma$. Observe that this is consistent with the chirality restriction. Using this and $\theta_0=0$,
we find that
 \bea
\hat\alpha(\sigma_0,\sigma_1,\dots, \sigma_4) \hat\alpha(\sigma_0,\bar\sigma_1,\dots, \bar\sigma_4)=e^{2i\psi}=1
 \eea
 and so $\psi=n \pi$, $n\in\bZ$. To preserve 30 real supersymmetries, we have to impose 15 conditions over the complex numbers.
 But since
 $e^{i\psi}=\pm1$, if $\hat\alpha(\sigma_0,\sigma_1,\dots, \sigma_4)=1$, then
 $(\hat\alpha(\sigma_0,\sigma_1,\dots, \sigma_4))^*=\hat\alpha(\sigma_0,\bar\sigma_1,\dots, \bar\sigma_4)=1$.
 Therefore
one can impose an even number of conditions each time. As a consequence supersymmetry can reduce only mod 2 over the complex numbers
or mod 4 over the reals. This in particular excludes the existence of discrete quotients with $N=30$ supersymmetries.

It remains to see whether the lift of (\ref{nulla}) can preserve 30 supersymmetries. In this case, we have
\bea
\hat\a=\exp\big({1\over2} [(\Gamma_0+\Gamma_5) \Gamma_9+\theta_1 \Gamma_{16}+\theta_2 \Gamma_{27}
+\theta_3 \Gamma_{38}]+i\psi\big)~.
\eea
Observe that this can be rewritten as
\bea
\hat\a=\rho\,\, [1+{1\over2} (\Gamma_0+\Gamma_5) \Gamma_9]~,~~~~\rho=\exp\big({1\over2} [\theta_1 \Gamma_{16}+\theta_2 \Gamma_{27}
+\theta_3 \Gamma_{38}]+i\psi\big)~.
\eea
Now the invariance condition can be written as
\bea
\rho\,\e_-=\e_-~,~~~\rho\,\e_++\rho\,\Gamma_{09}\, \e_-=\e_+~,
\eea
where we have decomposed the spinors in the eigenspaces $V_-\oplus V_+$ of $\Gamma_{05}$ as $\Gamma_{05}\e_\pm=\pm\e_\pm$. To preserve 30 supersymmetries
at least 7 complex spinors in $V_-$ must satisfy the first equation for $\e_-$. Since $\Gamma_{09}$ is invertible this would imply that
the second invariance equation cannot be satisfied on an at least  seven-dimensional
subspace of $V_+$. So there is no invariant complex  15-dimensional subspace in $V_-\oplus V_+$ which is required
to preserve 30 supersymmetries. Combining this with the result in the previous case, one concludes that
 there are  no  quotients of flat space that can  preserve
30 supersymmetries.

\subsection{Discrete quotients of $AdS_5\times S^5$}

The isometry group of this background is $SO(4,2)\times SO(6)$. Therefore one can choose $\alpha=e^{X+Y}$ where
$X\in \mathfrak{so}(4,2)$ and $Y\in \mathfrak{so}(6)$.  In addition, it can be arranged such that $Spin(4,2)\times Spin(6)$
acts on the Weyl representation of $Spin(9,1)$ as $\Delta^-_{Spin(4,2)}\otimes \Delta^-_{Spin(6)}$, where $\Delta^-_{Spin(4,2)}$
 $\Delta^-_{Spin(6)}$ are the anti-chiral Weyl representations of $Spin(4,2)$ and $Spin(6)$, respectively.
 Therefore the lifted element $\hat\alpha$ of $\alpha$ can be written as
 \bea
 \hat\alpha=e^{X+Y+i\psi}~,
 \eea
 where $X$ and $Y$ are Clifford algebra elements and
  $\psi$ is an additional  angle because of the $Spin_c(9,1)$ nature of the IIB spinors.

 There is a unique normal form for $Y$ up to a $Spin(6)$ conjugation which we can take to be
 \bea
 Y={1\over2}(\theta_1 \gamma_{12} +\theta_2 \gamma_{34}+\theta_3\gamma_{56})~,
 \eea
 where $\theta_1, \theta_2$ and $\theta_3$ are $SO(6)$ rotation angles, and $\gamma_i$ are $Spin(6)$ gamma matrices.
 Moreover $\Delta^-_{Spin(6)}$ can be decomposed in four complex one-dimensional spaces in which case one has that
 \bea
 Y={i\over2}(\sigma_1\theta_1  +\sigma_2\theta_2 +\sigma_3\theta_3)~,
 \eea
 where $\sigma_1\sigma_2\sigma_3=1$, $\sigma_i=\pm1$, due to the chirality restriction.

There are 25 possible normal forms for $X$ up to $SO(4,2)$ conjugations. These have be tabulated in \cite{simona} and we shall
not repeat them here. As a consequence, we have to investigate 25 cases to see whether there are quotients of $AdS_5\times S^5$
that preserve 30 supersymmetries. In what follows, we shall use the numbering of cases as in \cite{simona} but we have made some adjustments
in the notation because of our different spinor conventions.

\subsubsection{Cases 1, 2, 4, 10, 11, 12, 16, 24 and 25}

In case 24, the normal form for $X$ can be taken as
\bea
X={1\over2}(\zeta_1 \tilde \gamma_{05}+ \zeta_2 \tilde\gamma_{12} +\zeta_3\tilde\gamma_{34})~,
\eea
where $0$ and $5$ are the time-like directions and the rest are spacelike,  $\tilde\gamma$ are the gamma matrices
of $Spin(4,2)$ and $\zeta_i$ are angles. Decomposing $\Delta^-_{Spin(4,2)}$ in one-dimensional complex representations
we get that
\bea
X={i\over2}(s_1 \zeta_1 + s_2\zeta_2 +s_3 \zeta_3)~,
\eea
where $s_1s_2 s_3=1$ because of the chirality condition and $s_a=\pm1$.  Therefore the lifted element $\hat\alpha$ of
$\alpha$ is
\bea
\hat\alpha (s_1, s_2, \sigma_1, \sigma_2)= e^{{i\over2}(\sum_a s_a \zeta_a+\sum_i \sigma_i \theta_i)+i\psi}~.
\la{rot}
\eea
To preserve 30 supersymmetries $\hat\alpha (s, \sigma)=1$ for 15 out of 16 choices of signs for $s_a$ and $\sigma_i$
subject to the chirality conditions $s_1s_2 s_3=1$ and $\sigma_1\sigma_2\sigma_3=1$.  Without loss of generality let us assume
that $\hat\alpha (s, \sigma)=1$ unless when $\sigma_1=\sigma_2=s_1=s_2=-1$ for which we take
$\hat\alpha (-1, -1, -1, -1)\not=1$. Since $\hat\alpha (-1, -1, 1, 1)=\hat\a(1,1,1,1)=1$, then
\bea
(\hat\alpha (-1, -1, 1, 1))^* \hat\a(1,1,1,1)=e^{-i\zeta_1-i\zeta_2}=1~.
\eea
Then observe that
\bea
e^{-i\zeta_1-i\zeta_2} \hat\a(1,1,-1,-1)=\hat\alpha (-1, -1, -1, -1)=1~,
\eea
which is a contradiction. Therefore if one assumes that $\hat\a$ preserves  30 supersymmetries, then  one can show that
it preserves 32. So there are no such $N=30$ supersymmetric quotients of $AdS_5\times S^5$.

Before we proceed to other cases, notice that the same conclusion  holds if one of the angles $\zeta$ and/or one of the angles
$\theta$ vanish. This can be shown in exactly the same way as the general case above. In addition, if either  two or more angles
$\zeta$ vanish or two or more  angles $\theta$ vanish, then the decomposition of the Weyl representation of $Spin(9,1)$
with respect to $X+Y$
will be in subspaces of complex dimension more than one. Consequently, the invariant subspaces will have dimension either 32
and all supersymmetry will be preserved
or always less than 30. Therefore one concludes that there are no $N=30$ quotients even if one or more angles $\zeta, \theta$ vanish.

In the case 25 of \cite{simona}, the normal form of $X$ give rise to
\bea
X=\zeta_1 \tilde\gamma_{01}+\zeta_2 \tilde\gamma_{52}+ \zeta_3 \tilde\gamma_{34}~,
\eea
which after decomposing the Weyl representation in one-dimensional complex subspaces one gets
\bea
\hat\a(s_1, s_2, \sigma_1, \sigma_2)=e^{{1\over2}(s_1\zeta_1+s_2\zeta_2+i s_3 \zeta_3+ i \sum_i \sigma_i \theta_i)+i\psi}~,
\eea
where the signs $s$ and $\sigma$ obey the chirality conditions as in the previous case. In this case $\zeta_1$ and $\zeta_2$ are
boosts. If for some signs $\hat\a(s_1, s_2, \sigma_1, \sigma_2)=1$, then $(\hat\a(s_1, s_2, \sigma_1, \sigma_2))^*=1$, which
implies that
\bea
e^{s_1\zeta_1+s_2\zeta_2}=1~.
\eea
There are four possible uncorrelated  choices for the signs $s_1$ and $s_2$. To preserve $N=30$ supersymmetry for three
of these choices the above condition must hold. Without loss of generality one can take
\bea
e^{\zeta_1+\zeta_2}=e^{\zeta_1-\zeta_2}=1~.
\eea
This in turn gives $\zeta_1=\zeta_2=0$.  Consequently this  reduces to (\ref{rot}) with two vanishing angles.
As we have shown such quotients do not preserve 30 supersymmetries.  The same conclusion holds if one or more of the boosts
or rotation angles vanishes. Consequently, one can also conclude that the normal forms of the cases
1,2,4,10,11,12 and 16 \cite{simona} do not give quotients which preserve 30 supersymmetries.

\subsubsection{Cases 3, 5, 14, 15 and 17}

In  case 14, the lifted element is
\bea
\hat\a=\rho\, e^{{1\over2}\big((\tilde \gamma_0+\tilde \gamma_1) \tilde\gamma_5+\zeta \tilde\gamma_{23}\big)}
= \rho e^{{1\over2}\zeta \tilde\gamma_{23}} (1+A)~,
\eea
where $\rho\in Spin_c(6)$ and $A$ is a nilpotent generator, $A^2=0$. Decompose $\Delta^-_{Spin(4,2)}\otimes \Delta^-_{Spin(6)}=V_+\oplus V_-$
as $\tilde\gamma_{01}\epsilon_{\pm}=\pm \epsilon_{\pm}$. Then the invariance condition can be written as
\bea
\rho\, e^{{1\over2}\zeta \tilde\gamma_{23}} \epsilon_-&=&\epsilon_- ~,
\cr
\rho\, e^{{1\over2}\zeta \tilde\gamma_{23}} (\epsilon_++ \tilde \gamma_{05}\epsilon_-)&=&\epsilon_+ ~.
\eea
To preserve 30 supersymmetries, the first condition must be satisfied on an at least  seven-dimensional complex subspace $W_-$ of $V_-$.
In turn this implies that an at least seven-dimensional subspace $W_+$ of $V_+$ is also invariant. Thus if $\e_+\in W_+$, one concludes
that $\tilde \gamma_{50}\epsilon_-=0$, and since
$\tilde\gamma_{50}$ is invertible, $\e_-=0$, i.e.~the spinors in $W_-$ are not invariant.
 Therefore
such quotients  cannot preserve 30 supersymmetries. In fact one can show that $\hat\alpha$ preserves at most 16 supersymmetries.

The proof for cases 15 and 17 is similar. In addition,  3 and 5 are special cases. In all these cases, $N=30$ quotients
can be excluded.

\subsubsection{Case 7 and 19}

Let us begin with case 19. The lifted element can be written as
\bea
\hat\a=\rho e^{{1\over2}\varphi \tilde\gamma_{34}} e^{A+\zeta B}~,
\eea
where $\rho\in Spin_c(6)$ and
\bea
A={1\over2} (\tilde\gamma_5+\tilde\gamma_1) (\tilde\gamma_0+\tilde\gamma_2)~,
\cr
B={1\over2}( \tilde\gamma_{02}-\tilde \gamma_{51})~.
\eea
It is clear that the element generated by $\tilde\gamma_{34}$ commutes with all the other and
\bea
AB=BA=0~,~~~A^2=0~,~~~B^2=P_-~,~~~B^3=B~,~~~
\eea
where $P_\pm={1\over2} (1\pm\tilde\gamma_{0251})$. Using these, one finds that
\bea
e^{A+\zeta B}=(1+A)[P_++\cosh\zeta P_-+\sinh\zeta B]~.
\eea
Decomposing $\Delta^-_{Spin(4,2)}\otimes \Delta^-_{Spin(6)}=V_{++}\oplus V_{+-}\oplus V_{-+}\oplus V_{--}$
according to the commuting projections constructed from $\tilde\gamma_{51}$ and $\tilde\gamma_{02}$, one finds that the
invariance equation can be written as
\bea
\rho\, e^{{1\over2}\varphi \tilde\gamma_{34}} (\epsilon_{++}-2 \tilde\gamma_{05} \epsilon_{--})&=&\epsilon_{++}~,
\cr
\rho\, e^{{1\over2}\varphi \tilde\gamma_{34}}[\cosh\zeta \epsilon_{+-}+\sinh\zeta \epsilon_{+-}]&=&\epsilon_{+-}~,
\cr
\rho\, e^{{1\over2}\varphi \tilde\gamma_{34}} [\cosh\zeta\epsilon_{-+}-\sinh\zeta \epsilon_{-+}]&=&\epsilon_{-+}~,
\cr
\rho\, e^{{1\over2}\varphi \tilde\gamma_{34}} \epsilon_{--}&=&\epsilon_{--}~.
\eea
To obtain backgrounds with 30 supersymmetries, the last equation should have at least three complex independent solutions
$\epsilon_{--}$. This means
that there must exist angles $\theta, \psi$ and $\varphi$ such that $\rho\, e^{{1\over2}\varphi \tilde\gamma_{34}}=1$
for some selection of $\sigma$ signs. Substituting
this into the first equation, since the kernel of $\tilde\gamma_{05}$ is trivial, consistency requires that $\epsilon_{--}=0$.
Thus such solutions break more than 30 supersymmetries.  In addition, case 7 can be treated in a similar way.

\subsubsection{Cases 6, 8, 20 and 21}

The lifted element in case 20 can be written as
\bea
\hat\a=\rho e^{{1\over2}\varphi \tilde\gamma_{34}} e^{A+\zeta B}~,
\eea
where $\rho\in Spin_c(6)$ and
\bea
A={1\over2} (\tilde\gamma_5+\tilde\gamma_1) (\tilde\gamma_0+\tilde\gamma_2)~,
\cr
B={1\over2}( \tilde\gamma_{05}+\tilde \gamma_{12})~.
\eea
Next observe that
\bea
A^2=0~,~~~AB=BA~,~~~B^2=-P_-~,~~~B^3=-B~,~~~P_\pm={1\over2}(1\pm \tilde\gamma_{0512})~.
\eea
Using these, it is straightforward to show that
\bea
e^{A+\zeta B}=(1+A) [P_++\cos\zeta P_-+\sin\zeta B]~.
\eea
The rest of the analysis to exclude quotients which preserve 30 supersymmetries is similar to that of case 19 above.
In addition, cases 6,  8 and 21 can be treated in a similar way. All these cases do
not give quotients with 30 supersymmetries.

\subsubsection{Cases 9 and 22}

The lifted element in case 22 is
\bea
\hat\a= \rho\, e^{{1\over2}\varphi \tilde\gamma_{34}} e^{\zeta A+ \lambda B}~,
\eea
where
\bea
A={1\over2} (\tilde\gamma_{05}-\tilde\gamma_{12})~,~~~B={1\over2} (\tilde \gamma_{02}-\tilde \gamma_{51})~.
\eea
Observe that
\bea
AB=BA=0~,~~~A^2=-P_+~,~~~A^3=-A~,~~~B^2=P_-~,~~~B^3=B~,
\eea
where $P_{\pm}={1\over2} (1\pm \tilde\gamma_{0512})$. Using these we find that
\bea
e^{\zeta A+ \lambda B}&=&( P_-+\cos\zeta\, P_++\sin\zeta\, A) (P_+ +\cosh\lambda\, P_-+\sinh\lambda\, B)
\cr
&=&\cosh \lambda\, P_-+\cos\zeta\, P_++
\sinh\lambda\, B+\sin\zeta\, A~.
\eea
Decompose $\Delta^-_{Spin(4,2)}\otimes \Delta^-_{Spin(6)}=V_+\oplus V_-$ using  the projectors constructed
from $\tilde\gamma_{0512}$. Observing
that $B\epsilon_+=A\epsilon_-=0$, one can write the invariance equation as
\bea
\rho\, e^{{1\over2}\varphi \tilde\gamma_{34}} [ \cos\zeta \, \epsilon_++\sin\zeta \tilde\gamma_{05} \epsilon_+
+\cosh\lambda \epsilon_-+ \sinh\lambda  \tilde\gamma_{02}\epsilon_-]=\epsilon_++\epsilon_-\,.
\eea
Since $\tilde\gamma_{05}$ and $\tilde\gamma_{02}$ commute with the projectors constructed from $\tilde\gamma_{0512}$, one can rewrite
the invariance equations as
\bea
\rho\, e^{{1\over2}\varphi \tilde\gamma_{34}} \, e^{\zeta\tilde\gamma_{05}}\epsilon_+=\epsilon_+ ~,
\cr
\rho\, e^{{1\over2}\varphi \tilde\gamma_{34}}\, e^{\lambda \tilde\gamma_{02}}\epsilon_-=\epsilon_- ~.
\eea
The above invariance conditions can be simplified somewhat by observing that the $Spin(4,2)$ chirality
condition on the spinors together with the projections constructed from $\tilde\gamma_{0512}$
imply that $\tilde \gamma_{34} \epsilon_\pm=\mp i\epsilon_\pm$. To preserve  30 supersymmetries either $V_+$ or $V_-$ must
have a seven-dimensional invariant subspace.
Using a similar argument to the one we have presented in cases 24 and 25, one can easily show that if $V_+$ has a seven-dimensional
invariant subspace, then all of $V_+$ is invariant, and similarly for $V_-$. Therefore there are no such quotients with
30 supersymmetries. Case 9 can be analyzed in a similar way.

\subsubsection{Case 13}

The lifted element in this case is
\bea
\hat\a=\rho\, e^A~,
\eea
where
\bea
A={{1\over2} (\tilde \gamma_{05}+\tilde \gamma_{01}+\tilde \gamma_{03}-\tilde \gamma_{52}-\tilde \gamma_{12}-\tilde \gamma_{23}})~.
\eea
Observe that
\bea
A^2=-\tilde\gamma_{023} (\tilde\gamma_1+\tilde\gamma_5)~,~~~A^3={1\over2} \tilde\gamma_{12} (1+\tilde\gamma_{02}) (1+\tilde\gamma_{15})~.
\eea
Decomposing the spinors using the projectors constructed by $\tilde\gamma_{15}$ and $\tilde\gamma_{02}$, one finds that the invariance equation
can be decomposed as
\bea
\rho\, \e_{++}&=&\e_{++}~,
\cr
\rho\, (\e_{+-}+\tilde\gamma_{03}\e_{++})&=&\e_{+-}~,
\cr
\rho\, (\e_{-+}+2\tilde \gamma_{01} \e_{+-}+\tilde \gamma_{13}\e_{++})&=&\e_{-+}~,
\cr
\rho\, (\e_{--}+\tilde \gamma_{03}\e_{-+}-\tilde\gamma_{13}\e_{+-}+{1\over3} \tilde \gamma_{12}\e_{++})&=&\e_{--}~.
\eea
It is straightforward from these to argue that there are no so such quotients which preserve 30 supersymmetries.

\subsubsection{Cases 18 and 23}

The lifted element for case 18 is
\bea
\hat\a=\rho\, e^{\zeta A+B}~,
\eea
where
\bea
A={1\over2} (\mp \tilde \gamma_{05}+\tilde \gamma_{12}+\tilde \gamma_{34})~,~~~B={1\over2}(\tilde \gamma_{03}-\tilde \gamma_{13}
\pm \tilde\gamma_{54}-\tilde\gamma_{24})~.
\la{sig}
\eea
Next observe that
\bea
[A,B]=0~,~~~B^3=0~.
\eea
Using these and without loss of generality choosing one of the signs in (\ref{sig}), one finds that the equation
for invariance can be written as
\bea
\rho\, e^{\zeta A} [\e_{++}+\e_{--}+\tilde\gamma_{03} \e_{+-}+\tilde \gamma_{54}  \e_{-+}+\tilde\gamma_{0543}\e_{++}]&=&\e_{--}+\e_{++}~,
\cr
\rho\, e^{\zeta A} [\e_{+-}+\e_{-+}+\tilde\gamma_{03} \e_{++}+\tilde \gamma_{54}  \e_{++}]&=&\e_{+-}+\e_{-+}~,
\la{xxy}
\eea
where we have decomposed $\Delta^-_{Spin(4,2)}\otimes \Delta^-_{Spin(6)}=V_{++}\oplus V_{-+}\oplus V_{+-}\oplus V_{--}$ with respect to the projectors
constructed from $\tilde\gamma_{01}$ and $\tilde\gamma_{52}$,
and use the property of $A$ to commute with $\tilde\gamma_{0152}$. In addition, using the property of $A$ to commute with
the projectors ${1\over4} (1\pm\tilde\gamma_{01}) (1\pm \tilde\gamma_{52})$, with the signs correlated, the first equation
in (\ref{xxy}) can be decomposed further as
\bea
\rho\, e^{\zeta A} \e_{++}&=&\e_{++}~,
\cr
\rho\, e^{\zeta A} [\e_{--}+\tilde\gamma_{03} \e_{+-}+\tilde \gamma_{54}  \e_{-+}+\tilde\gamma_{0543}\e_{++}]&=&\e_{--}~.
\eea
To preserve 30 supersymmetries, the first equation above has to have at least three solutions. On these solutions,
one can show that $\rho e^{{i\over2}\zeta}=1$. On the three dimensional  eigenspace in  $V_{--}$ of $\rho\, e^{\zeta A}$ with the same eigenvalues
consistency requires that
\bea
\tilde\gamma_{03} \e_{+-}+\tilde \gamma_{54}  \e_{-+}+\tilde\gamma_{0543}\e_{++}=0~.
\eea
This condition can be solved to express at least three complex components of $\e$ in terms of the remaining 13 components. Thus
there are not 15 independent complex solutions to the invariance condition, and so such quotients cannot preserve 30 supersymmetries.
The case 23 can be treated in a similar way.

\subsection{Discrete quotients of plane wave}

The isometry superalgebra\footnote{We have not included the anti-commutator of the odd generators $Q$ because
it is not used in the analysis.} of the maximally supersymmetric plane wave \cite{mspw} is
\bea
&&[e_-, e_i]=e^*_i~,~~~[e_-, e^*_i]=-4\lambda^2 e_i~,~~~[e^*_i, e_j]=-4\lambda^2 \delta_{ij} e_+ ~,
\cr
&&[M_{ij}, e_k]=-\delta_{ik} e_j+\delta_{jk} e_i~,~~~[M_{ij}, e^*_k]=-\delta_{ik} e^*_j+\delta_{jk} e^*_i~,~~i,j=1,2,3,4\,
\mathrm{and}\, 6,7,8,9
\cr
&&[e_+, Q]=0~,~~~[e_-, Q]=i \lambda (I+J) Q ~,
\cr
&&[e_i, Q]=-i\lambda I \Gamma_i \Gamma_+Q~,~~~ [e^*_i, Q]=-2\lambda^2 I \Gamma_i \Gamma_+Q~,~~~i=1,2,3,4
\cr
&&[e_i, Q]=-i\lambda J \Gamma_i \Gamma_+Q~,~~~[e^*_i, Q]=-2\lambda^2 J \Gamma_i \Gamma_+Q~,~~~i=6,7,8,9
\cr
&&[M_{ij}, Q]={1\over2} \Gamma_{ij} Q~,~~~I=\Gamma_{1234}~,~~~J=\Gamma_{6789}~,
\la{liealg}
\eea
where $\lambda$ is a real parameter.
It can be read off from (\ref{liealg})  that the isometry algebra of the maximally supersymmetric plane wave is $\mathfrak{so}(4)\oplus
\mathfrak{so}(4)\oplus_s \mathfrak{t}$, where $\mathfrak{t}= \mathfrak{so}(2)\oplus_s\mathfrak{h}_{17}$ and $\mathfrak{h}_{17}$
is a Heisenberg algebra. The most general element of the isometry Lie algebra is
\bea
X=u^+e_++v^- e_-+ v^i e_i+ w^i e^*_i+ {1\over2} \theta^{ij} M_{ij}~,
\eea
where the indices $i$ and $i,j$ are restricted as in  (\ref{liealg}).
Up to a conjugation, $X$ can be brought to either
\bea
X=u^+ e_++v^- e_-+\sum_{n=0, n\not=2}^4w^{2n+1}e_{2n+1}^*+\theta^1 M_{12}+\theta^2 M_{34}+\theta^3 M_{67}+\theta^4 M_{89}
\eea
if $v^-\not=0$, or
\bea
X=u^+ e_++\sum_{i=1, i\not=5}^9 v^i e_i+\sum_{n=0, n\not=2}^4w^{2n+1}e_{2n+1}^*+\theta^1 M_{12}+\theta^2 M_{34}+\theta^3 M_{67}+\theta^4 M_{89}
\eea
if $v^-=0$. The action of the isometries on the Killing spinors can be read off from the commutators of the generators
of the isometries with those of super-translations. In particular a lifted element is
\bea
\hat\a=e^{A+B}~,
\eea
where
\bea
A&=&i v^-  \lambda (I+J)+{1\over2} ( \theta^1\Gamma_{12}+\theta^2\Gamma_{34}+\theta^3 \Gamma_{67}+\theta^4\Gamma_{89})+i\psi ~,
\cr
B&=&- \lambda [ I\sum_{i=1}^4 \Gamma_i (i v^i+2\lambda w^i)+ J\sum_{i=6}^9 \Gamma_i (i v^i+2\lambda w^i)] \Gamma_+~.
\eea
The lifted generator $\hat\a$ has  been partially adapted to the normal forms of $X$
 but the expression above will suffice for the analysis that follows.
The Killing spinors are invariant along $e_+$ translations and so any identification along this direction preserves
all supersymmetry.
 Writing $\e=\e_++\e_-$ with $\Gamma_+\e_+=0$,
we find that the invariance condition can be written as
\bea
e^A\e_-&=&\e_- ~,
\cr
e^A(\e_++ \Gamma_+ \b \e_-)&=&\e_+ ~,
\eea
where $\b$ is a linear map that can be determined. Let us start by examining the first equation. The chirality of IIB spinors
together with the lightcone projection implies that $(I+J)\e_-=0$. Therefore only the rotation part of $e^A$ acts on $\e_-$.
Thus one has
\bea
e^{{i\over2} \sum_{i=1}^4 \sigma_i \theta_i+i\psi}\e_-=\e_-~, ~~~~\sigma_1\sigma_2\sigma_3\sigma_4=-1~.
\eea
The restriction on the $\sigma$ is due to the chirality condition on the spinors. There are 8 choices of signs giving
rise to 8 independent conditions. $N=30$ supersymmetry requires
that   at least 7
 conditions must hold. However one can show that if 7 conditions hold, then they imply the 8th.
 Moreover $\theta_i=2\pi n_i$ and $\psi=n_0 \pi$, where $n_0, n_i\in \bZ$. These angles are associated with the
  identity rotation  which
lifts to the identity element, so in what  follows we shall set $\theta_i=\psi=0$. However observe that the invariance condition
on $\e_-$ does not restrict $v^-$.

Next let us turn to the second equation and consider the case $v^-=0$.  Then
 to preserve 30 supersymmetries, the kernel of
$\b$ should have complex dimension 7. It turns out that
\bea
\b\e_-=\lambda [I\sum_{i=1, i\not=5}^9 \Gamma_i (i v^i+2\lambda w^i)]\e_- ~.
\eea
So there is a non-trivial kernel iff
\bea
- v^2+4\lambda^2  w^2-4i\lambda  v\cdot  w=0~,
\eea
which in turn implies that $ v\cdot  w=0$ and $ v^2=4\lambda^2  w^2$. However in such
a case the kernel has dimension 4 or 8. The latter occurs if $ v= w=0$. Thus there are no $N=30$ quotients for $v^-=0$.

Next let us consider the case where $v^-\not=0$. In such a case the $e_-$ generator acts non-trivially on $\e_+$. To
continue observe that $\hat\a$ factorizes as
\bea
\hat\a=e^{i v^-  \lambda I - \lambda  I\sum_{i=1}^4 \Gamma_i (i v^i+2\lambda w^i) \Gamma_+}\,\, e^{i v^- \lambda J-\lambda
J\sum_{i=6}^9 \Gamma_i (i v^i+2\lambda w^i) \Gamma_+}~.
\eea
Using that $I$ and $I\Gamma_i\Gamma_+$ anti-commute and the latter is nilpotent, and similarly for $J$ and  $J\Gamma_i\Gamma_+$,
and after some computation, one finds that
\bea
e^{2i\lambda v^- I}\e_++\Gamma_+ {\sin(\lambda v^-) e^{i\lambda v^-I}\over \lambda v^-} \sum_{i=1, i\not=5}^9[ i\lambda v^i+2\lambda^2 w^i]
 I\Gamma_i \e_-=\e_+~.
\eea
Thus one has that
\bea
\b={\sin(\lambda v^-) e^{i\lambda v^-I}\over \lambda v^-} \sum_{i=1, i\not=5}^9[ i\lambda v^i+2\lambda^2 w^i]
I \Gamma_i~.
\eea
As in the case with $v^-=0$, we have to investigate the kernel of $\b$.
 If $\lambda v^-=n \pi$, $n\in\bZ-\{0\}$, then all supersymmetry is preserved. As it can be seen,
it is remarkable that the Killing spinors
in \cite{mspw} are periodic in $v^-$ with precisely this period. If $\lambda v^-\not=n \pi$, then $\beta$ has a non-trivial kernel iff
$v^2=4\lambda^2 w^2$ and $v\cdot w=0$. As in the case with $v^-=0$, one concludes that the kernel has dimension either 4
or 8. Thus such quotients do not preserve 30 supersymmetries.

\newsection{Concluding remarks}

We have shown that all $N>28$ supersymmetric IIB backgrounds are maximally supersymmetric. The proof relies
on the property that these backgrounds have vanishing one-form and  three-form fluxes, $P=G=0$, which arises as consequence
of the homogeneity of $N>24$ backgrounds and
the algebraic Killing spinor equation of IIB supergravity. In addition,   the supercovariant curvature vanishes
subject to the field equations and the Bianchi identities of the theory. Therefore all $N>28$ supersymmetric IIB backgrounds
are locally maximally supersymmetric. Finally,
$28<N<32$ backgrounds cannot be constructed as discrete quotients  of maximally supersymmetric ones.

It is natural to ask whether it is possible to extend the above results to other  near maximal backgrounds with $N\leq 28$.
This does not seem straightforward.  In particular, it is known that there are plane wave backgrounds
 with 28 supersymmetries \cite{bena, michelson}. Significantly, these backgrounds have non-vanishing three-form flux, $G\not=0$.
Thus apart from the maximally supersymmetric case,   $7/8$ is the highest fraction of supersymmetry that  IIB backgrounds preserve.

 The existence of backgrounds with 28 supersymmetries does not necessarily imply that there are supersymmetric
 backgrounds for all $N < 28$. It may be that backgrounds with a particular number of supersymmetries can be excluded. Such cases
 will exhibit supersymmetry enhancement similar to that we have shown for backgrounds with $N>28$.
 It would be of interest to classify all IIB backgrounds with 28 supersymmetries as the first near maximal
 case that has solutions which do not have maximal supersymmetry. This may be possible
  using the homogeneity of these backgrounds.

 Our results can be extended to investigate nearly maximally supersymmetric IIA backgrounds.
 This is because of the similarities between the Killing spinor equations of IIA and IIB supergravities; in particular
 both have an algebraic Killing spinor equation. In fact, it appears  that the nearly maximally supersymmetric
 solutions of IIA supergravity are more restricted than those of IIB. In particular,
 there is a unique maximally supersymmetric IIA solution, the Minkowski
 spacetime, and the $N=31$ IIA backgrounds are maximally supersymmetric.
 The $N=30$ IIA backgrounds can be investigated
 in a way similar to those of IIB by appropriately modifying the IIB complex linearity argument for the IIA dilatino Killing spinor equation
 and showing that the supercovariant curvature vanishes.

 In eleven-dimensions, the investigation of nearly maximally supersymmetric backgrounds is more involved. This
 is because eleven-dimensional supergravity  does not have an algebraic Killing spinor equation.
 So an extension of our results to eleven-dimensions
 depends crucially on the properties of the gravitino Killing spinor equation. Nevertheless, it would be
 of interest to see whether the results of \cite{N=31-M} can be extended to backgrounds with less than 31 supersymmetries.

\vskip 0.5cm
\noindent{\bf Acknowledgements} \vskip 0.1cm

GP is partially supported by the PPARC grant PP/C507145/1
and the EU grant MRTN-CT-2004-512194. The work of DR has been supported by the European EC-RTN project
MRTN-CT-2004-005104, MCYT FPA 2004-04582-C02-01 and CIRIT GC
2005SGR-00564. The work of U.G.~is funded by the Swedish Research Council.

\vskip 0.5cm

\
\setcounter{section}{0}

\end{document}